\documentclass[prb,reprint,twocolumn,amssymb,superscriptaddress]{revtex4-2}

\usepackage{xcolor}

\usepackage{bm}
\usepackage{amsmath}  
\usepackage{amsfonts}
\usepackage{graphicx}
\usepackage{hyperref}
\usepackage{comment}
\usepackage{soul}
\hypersetup{
	colorlinks=true,
	citecolor=blue,
	linkcolor=red,
	urlcolor = blue
}

\newcommand{\xvec}{\bm{x}}
\newcommand{\pvec}{\bm{p}}
\newcommand{\qvec}{\bm{q}}
\newcommand{\Hhat}{\hat{H}}

\newcommand{\grad}{\nabla}

\begin{document}
\title{Nonadiabatic Origin of Quantum-Metric Effects via Momentum-Space Metric Tensor}
\author{Yafei Ren}
\affiliation{Department of Physics and Astronomy, University of Delaware, Newark, DE 19716, USA}

\begin{abstract}
We reveal a fundamental geometric structure of momentum space arising from the nonadiabatic evolution of Bloch electrons. By extending semiclassical wave packet theory to incorporate nonadiabatic effects, we introduce a momentum-space metric tensor---the nonadiabatic metric. This metric gives rise to two velocity corrections, dubbed geometric and geodesic velocities, providing a unified and intuitive framework for understanding nonlinear and nonadiabatic transport phenomena beyond Berry phase effects. {The geometric velocity is related to the nonadiabatic metric itself, whereas the geodesic velocity is a Christoffel symbol of the nonadiabatic metric. As the nonadiabatic metric is related to the energy-gap renormalized quantum metric, it unifies the broad quantum metric effects in electronic responses.}
When the nonadiabatic metric is constant, it reduces to an effective mass, modifying flat-band electron dynamics in confining potentials. In a flat Chern band with harmonic attractive interactions, the two-body wave functions mirror the Landau-level wave functions on a torus. Furthermore, we show that the nonadiabatic metric endows momentum space with a curved geometry, recasting wave packet dynamics as forced geodesic motion. 
\end{abstract}

\maketitle

\section{Introduction}
{The adiabatic approximation is a cornerstone of condensed matter physics. The discovery of the Berry phase in adiabatic evolution~\cite{berry1984quantal} has profoundly shaped the study of topological phases of matter and provided a unified description of electromagnetic response properties, including polarization and orbital magnetization~\cite{xiao2010berry}.

The Berry curvature arises as the imaginary part of the quantum geometric tensor~\cite{provost1980riemannian}, which characterizes the geometry of quantum states in parameter space. The real part of the tensor defines the quantum metric, which encodes the distance between nearby quantum states. Motivated by early seminal works~\cite{parameswaran2013fractional, peotta2015superfluidity} and by the expectation that this complementary geometric quantity may play a role comparable to that of the Berry curvature~\cite{torma2023essay}, recent years have witnessed rapidly growing interest in identifying physical consequences of the quantum metric. A wide range of phenomena have been discussed under the umbrella of quantum-metric-related effects, including Wannier function spread~\cite{marzari1997maximally}, nonlinear transport~\cite{yu2024quantum, liu2025quantum, verma2025quantum, neupert2013measuring, gao2014field, morimoto2016topological, ozawa2018steady, gao2019nonreciprocal, ahn2020low, ahn2022riemannian, nag2023third, ma2023anomalous, bouhon2023quantum, jankowski2024quantized, jankowski2025optical, avdoshkin2024multi, sala2024quantum}, orbital magnetic susceptibility~\cite{gao2015geometrical, piechon2016geometric}, massive particle dynamics in flat bands~\cite{julku2016geometric, liang2017band, hu2019geometric, xie2020topology, julku2020superfluid, rossi2021quantum, torma2022superconductivity, tian2023evidence}, and stabilizing fractional Chern insulators~\cite{wang2021exact, ledwith2023vortexability, estienne2023ideal, liu2024theory}.

Despite this progress, the rapidly expanding literature has also led to conceptual ambiguities. Accumulated evidence indicates that electronic dynamics and responses are not governed directly by the quantum metric itself~\cite{yu2024quantum, liu2025quantum, verma2025quantum}. Rather, they are controlled by derived quantities involving energy-gap-weighted interband matrix elements~\cite{yu2024quantum, liu2025quantum, verma2025quantum, neupert2013measuring, gao2014field, morimoto2016topological, ozawa2018steady, gao2019nonreciprocal, ahn2020low, ahn2022riemannian, nag2023third, ma2023anomalous, bouhon2023quantum, jankowski2024quantized, jankowski2025optical, avdoshkin2024multi, sala2024quantum, gao2015geometrical, piechon2016geometric, julku2016geometric, liang2017band, hu2019geometric, xie2020topology, julku2020superfluid, rossi2021quantum, torma2022superconductivity, tian2023evidence}. The same matrix elements appear in the band summation expression of the quantum metric. This observation highlights the need for a closer examination of the concepts involved in quantum-metric-related effects, to identify the fundamental quantities that control the dynamics and to organize the associated phenomena in a more refined manner.

From a dynamical perspective, an additional motivation for this work is the contrast between Berry-curvature-driven transport and quantum-metric-related phenomena. While the role of Berry curvature is well understood, being intrinsically tied to adiabatic evolution and entering electron dynamics directly through the anomalous velocity~\cite{xiao2010berry}, a comparably transparent dynamical picture for quantum-metric-related transport phenomena remains less developed, despite substantial recent progress summarized in those reviews~\cite{yu2024quantum, liu2025quantum, verma2025quantum}. Within the semiclassical wave-packet framework, Gao and co-workers extended the theory to higher-order responses to electric and magnetic fields~\cite{gao2014field, gao2015geometrical}.
While this formulation provides a compact description of nonlinear responses, the interband dynamics are encapsulated in the corrections of Berry connections. Since the original Berry connection is defined in the adiabatic limit, it becomes unclear whether these corrections are genuinely adiabatic in nature or instead originate from nonadiabatic interband coherence. 
This ambiguity is further underscored by the appearance of geometric objects such as the Christoffel symbols~\cite{smith2022momentum, jain2024anomalous, mehraeen2025quantum}. These symbols are conventionally associated with particle dynamics in a curved space defined by a metric tensor, whereas the metric tensor does not enter their formulation explicitly~\cite{gao2014field, gao2015geometrical}. These mismatches call for a more fundamental reexamination of the geometric structures governing electronic dynamics.

In this work, we address this issue by showing that it is not the quantum metric itself, but a closely related quantity dubbed the nonadiabatic metric that provides the unifying geometric framework for nonlinear and nonadiabatic transport phenomena. Unlike the conventional quantum metric, this metric is intrinsically tied to nonadiabatic evolution and enters electron dynamics directly as a momentum-space metric tensor. This metric tensor naturally leads to geodesic equations with Christoffel symbols. It thereby offers a transparent dynamical and geometric picture for quantum-metric-related transport phenomena, placing them on a footing comparable to Berry-curvature-driven transport.

We obtain these results by extending the adiabatic wave packet theory to the nonadiabatic regime by including interband contributions (see Fig.~\ref{fig:diabatic}). 
We begin by deriving the effective Lagrangian for the wave packet. 
We find that the leading-order nonadiabatic corrections beyond the adiabatic effects emerge as a metric tensor in momentum space, which we term the nonadiabatic metric. We show that this metric is related to but different from the quantum metric. Next, we elucidate the profound implications of the nonadiabatic metric. We first discuss its impact on the wave packet velocity, including a geometric velocity proportional to the metric itself times the time-derivative of the electric field, and a geodesic velocity proportional to the Christoffel symbol times the electric field squared. 
These velocities provide a unified understanding of nonlinear and nonadiabatic transport. 
Then we show explicitly that the nonadiabatic metric endows the Brillouin zone with the structure of a curved manifold, causing electron dynamics to resemble forced geodesic motion on this curved space. This gives rise to the Christoffel symbol naturally. We further study the quantum dynamics of Bloch electrons by requantizing the semiclassical wave-packet dynamics in momentum space, where the toroidal geometry of the Brillouin zone imposes periodic boundary conditions. When applied to flat Chern bands, this approach reveals the reemergence of massive particle dynamics where the nonadiabatic metric induces a correction on the bound-state spectrum and the bound-state wave functions are related to quantum Hall states on a torus.

\begin{figure}[t!]
\includegraphics[width=0.5\linewidth]{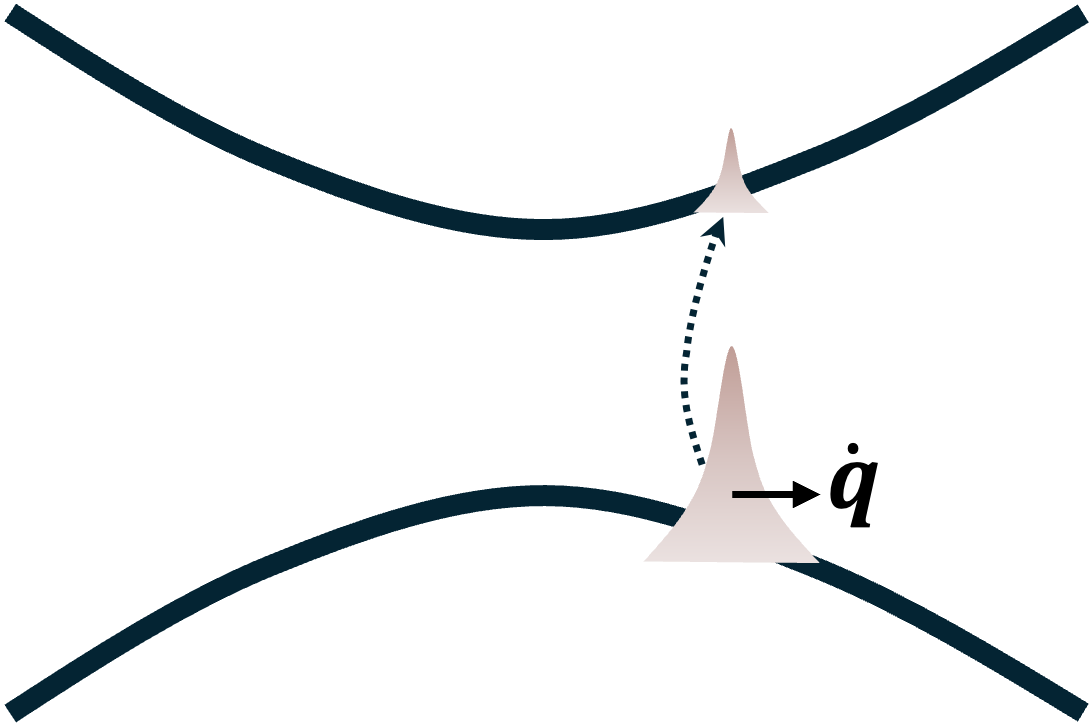}
\caption{Illustration of nonadiabatic time evolution of a wave packet in momentum space, where a gradual momentum shift induces interband mixings and nonadiabatic corrections.}
\label{fig:diabatic}
\end{figure}

Our results indicate that, in the dynamics of the Bloch electrons, the nonadiabatic metric is more fundamental than the quantum metric: it emerges directly from the Schr\"odinger equation as the leading correction beyond adiabatic evolution. It governs the dynamical behavior of Bloch electrons, providing a unified description of various transport phenomena. In contrast, the quantum geometric tensor is defined to describe the geometric properties of static wave functions on a parameter space~\cite{provost1980riemannian}, which are naturally tied to static properties or adiabatic processes.

In addition, nonadiabatic effects have been extensively studied in molecular dynamics and in quantum systems with internal degrees of freedom~\cite{jackiw1988three, kolodrubetz2017geometry, goldhaber2005newtonian, requist2016molecular, scherrer2017mass, littlejohn2023representation, nelson2020non, allahverdyan2009post, hetenyi2023fluctuations}. In those settings, the slow variables are typically externally prescribed parameters, such as atomic coordinates~\cite{jackiw1988three, kolodrubetz2017geometry, goldhaber2005newtonian, requist2016molecular, scherrer2017mass, littlejohn2023representation}. By contrast, in our semiclassical theory of Bloch electrons, the slow variables emerge as collective coordinates of trial wave functions constructed from Bloch states, which are critical for understanding the geometric organization of electronic responses in crystalline solids.


In the following, we present our formalism in Sec.\,\ref{sec:formalism}, deliver the physical manifestations of the nonadiabatic metric on electronic responses in Sec.\,\ref{sec:results}, and discuss the geometric effects of the nonadiabatic metric for electron dynamics in momentum space in Sec.\,\ref{sec:qspace}. 
We summarize our results in Sec.\,\ref{sec:summary}.
}

\section{Formalism: Nonadiabatic Wave Packet Theory} \label{sec:formalism}

\subsection{Wave Packet Ansatz}

We study the dynamics of a wave packet centered at \(\xvec_c\) under slowly varying scalar potential $V(\bm{x})$. Its spatial spread is much smaller than the perturbation scale, which enables us to consider an approximated Hamiltonian local to the wave packet. The Hamiltonian can be approximated by linearizing perturbations around the wave packet center:
\begin{align}
    \mathcal{H} \approx& \Hhat_c + \hat{H}_1, 
\end{align}
where \(\Hhat_c = \hat{H}(\hat{\xvec}, \hat{\pvec}) + V(\xvec_c)\) represents the local Hamiltonian with $V(\xvec_c)$ a constant and \(\Hhat\) the Hamiltonian operator of the unperturbed crystal. The corresponding Bloch bands and Bloch wave functions satisfying:
\begin{align}
    \Hhat |\psi_{n,\bm{q}}\rangle = E_{n}(\qvec) |\psi_{n,\qvec}\rangle,
\end{align}
$\hat{H}_1=(\hat{\xvec} - \xvec_c) \cdot \grad_{\xvec} V(\xvec_c) $ is the gradient correction. Here $\grad_{\xvec} V(\xvec_c)$ are constants and do not induce inter-band coupling~\cite{sundaram1999wave}.

We use the following wave packet ansatz~\cite{SM}:
\begin{align}
|\Psi(t)\rangle = \int d\bm{q} \, a(\boldsymbol{q}, t) |\tilde{\psi}_{\boldsymbol{q}}\rangle,    
\end{align}
where \( a(\boldsymbol{q}, t) = |a| e^{-i \gamma} \) is a narrow distribution in momentum space centered at \( \boldsymbol{q}_c = \int  \bm{q}|a|^2 d\bm{q} \), $|\tilde{\psi}_{\boldsymbol{q}}\rangle 
= e^{i \boldsymbol{q} \cdot \hat{\boldsymbol{x}}} |\tilde{u}_{\bm{q}}\rangle$ is a superposition of Bloch waves with quasimomentum $\bm{q}$, and
\begin{align}
|\tilde{u}_{\bm{q}}\rangle 
= \sum_n c_n(\boldsymbol{q}, t) |u_n(\boldsymbol{q})\rangle 
\end{align}
with \( |u_n(\boldsymbol{q})\rangle \) the periodic part of the Bloch wave function $|\psi_{n,\bm{q}}\rangle$, $c_n$ the coefficient, and \( \sum_n |c_n(\boldsymbol{q}, t)|^2 = 1 \). The wave packet center is at $\bm{x}_c = \langle\Psi|\hat{\bm{x}}|\Psi\rangle$.

Here we consider an isolated band labeled by index 0. 
We focus on the abelian case with \( |c_0| \approx 1 \) and $c_n \sim 0$ for $n \neq 0$. 
In contrast to the adiabatic wave packet dynamics developed by Sundaram and Niu~\cite{sundaram1999wave}, which constrains the wave packet on a single energy band, here we allow for interband contributions with $c_n \neq 0$ (see Fig. 1). These interband components $c_n$ are subsequently expressed as an asymptotic series in the rate of parameter evolution such as $\dot{\bm{q}}_c$. This contrasts with the higher-order expansion in Ref.~\cite{gao2014field}, where the wave packet is expanded as a series in the strength of external fields rather than in the rate of change of the underlying parameters.

\subsection{Wave Packet Dynamics and Effective Lagrangian}
The dynamics of the wave packet is obtained by minimizing the Dirac-Frenkel action (with $\hbar=1$)~\cite{dirac1930note, raab2000dirac}
\begin{align}
    \mathcal{S}
    &=\int dt \langle\Psi|i\frac{d}{dt}-\mathcal{H}|\Psi\rangle = \int dt L \\
    L&= - \dot{\bm{q}}_c \cdot \bm{x}_c - E_{c} + \Lambda_{[00]} \notag \\
&+ \sum_n c_n^* i \dot{c}_n + \sum_{n\neq 0} -c_n^* c_n \Delta_n + c_n^* J_n + c_n J_n^*
\end{align}
where $E_c = E_0(\bm{q}_c)+V(\bm{x}_c)$ with $E_n$ the eigenenergy of the $n$-th band and $V(\bm{x}_c)$ the slowly varying external potential. 
Here $\dot{c}_n=\frac{dc_n}{dt}$, $\Delta_n \equiv \omega_{[n0]} - (\Lambda_{[nn]}-\Lambda_{[00]})$, $\Lambda_{[mn]}=\dot{\bm{q}}_c\cdot \bm{A}_{[mn]}$, $J_n \equiv c_0 \Lambda_{[n0]}$, 
$\omega_{[mn]}=E_m-E_n$, and ${A}_{i,[mn]}=\langle u_m | i\partial_{{q}^i} | u_n \rangle|_{\bm{q}_c}$. The dynamical variables in the Lagrangian are $\bm{x}_c$, $\bm{q}_c$, $c_n$, and $c_n^*$. Detailed derivation of the above Lagrangian is shown in Sec. I of the Supplemental Material~\cite{SM}.

Varying $\mathcal{S}$ with respect to $c_n^*$ gives the saddle-point equation
\begin{align}
(i\partial_t-\Delta_n)c_n = -J_n\qquad n\neq0. \label{eq:cn-eom}
\end{align}
Let $G_n$ be the retarded Green’s operator of $i\partial_t-\Delta_n$, i.e.
$(i\partial_t-\Delta_n)G_n=\delta(t-t’)$.
Solving Eq.~\eqref{eq:cn-eom} obtains $c_n(t)=-\int G_n(t, t') J_n(t')dt'$. 
Substituting it back into the action, the terms including $i\dot c_n, \Delta_n, c_0\Lambda^{n0}$ cancel out. Expressing $c_0^* \Lambda^{0n}$ as $J_n^*(t)$, one yields 
\begin{align}
\mathcal{S}_{\rm eff}[\bm q_c,\bm x_c] 
&= \int dt L_0(\bm q_c,\bm x_c) \notag \\
&- \sum_{n\neq0}\int dt dt' J_n^*(t) G_n(t,t')J_n(t'), \label{eq:S-eff-nonlocal}
\end{align}
where $L_0(\bm q_c,\bm x_c) \equiv - \dot{\bm q}_c \cdot \bm{x}_c - E_{c} + \Lambda_{[00]}$ is the adiabatic Lagrangian of the wave packet.

When parameters $\bm q_c$ vary slowly, we expand the inverse operator:
\begin{align}
G_n 
&=(i\partial_t-\Delta_n)^{-1} \delta(t-t') \notag \\
&=\left[- \frac{1}{\Delta_n} +\mathcal O(\partial_t)\right] \delta(t-t'). \label{eq:Gn-exp}
\end{align}
Using $J_n=c_0\Lambda^{n0}$ and integrating by parts, equation \eqref{eq:S-eff-nonlocal} gives an effective Lagrangian to the leading order of the nonadiabatic correction
\begin{align}
L_{\rm eff} = L_0
+\sum_{n\neq0}\frac{J_n^*J_n}{\Delta_n}
+\mathcal O(\partial_t)
\label{eq:Leff-compact}.
\end{align}
By substituting $J_n=c_0 \dot{\bm{q}}_c\cdot \bm{A}_{[n0]}$ into the above equation and neglecting higher-order corrections by taking $|c_0|^2 \simeq 1$ and ignoring the Berry-phase corrections to the band gap, i.e., $ |\Lambda_{[nn]}-\Lambda_{[00]}| \ll |\omega_{[n0]}|$ and $\Delta_n\simeq \omega_{[n0]}$, we obtain our central result: the nonadiabatic effective Lagrangian 
\begin{align} \label{eq:Leff}
    L_{\rm eff} &= - \dot{\bm{q}}_c \cdot \bm{x}_c - E_c(\bm{q}_c,\bm{x}_c) + \dot{q}_c^i {A}_i + \frac{1}{2} 
    G_{ij} \dot{q}_c^i \dot{q}_c^j 
\end{align}
where the Einstein summation convention is used for Cartesian coordinate indices throughout the manuscript, while summation over the band indices is written explicitly. 
${A}_i$ is used as shorthand for ${A}_{i,[00]}$, capturing the Berry phase effects from adiabatic evolution of Bloch states.
We emphasize that this work focuses on a slowly varying scalar potential, resulting in a gradient correction $\hat{H}_1$ that is proportional to a constant times $\hat{\bm{x}}-\bm{x}_c$. This specific form of $\hat{H}_1$ does not induce inter-band coupling. However, in the presence of a magnetic field or inhomogeneities, such as those arising from spin textures, the nonadiabatic theory yields additional corrections to the energy $E_c$ and Berry connections $A_i$ as discussed in our follow-up work~\cite{ren2025nonadiabatic}.
From now on, we drop the subscript $c$, using $(\bm{x},\bm{q}, E)$ for $(\bm{x}_c,\bm{q}_c, E_c)$.

\subsection{Nonadiabatic Metric}

Our key contribution is the identification of a metric term beyond the Berry phase effects, characterized by the coefficients $G_{ij}$. $G_{ij}$ acts as a metric tensor in momentum space, which is termed the nonadiabatic metric, as this metric originates from nonadiabatic corrections to wave packet dynamics: $c_n \propto \dot{q}^i$ for $n\neq 0$, which reflect a finite occupation of other instantaneous eigenstates.
The matrix element reads
\begin{align} \label{eq:nonadiabaticMetric}
    G_{ij} 
    =& 2\text{Re}\sum_{m \neq 0} \frac{A_{i,[0m]} A_{j,[m0]}}{\omega_{[m0]}} .
\end{align}
The metric is a gauge-invariant symmetric tensor~\cite{nakahara2018geometry}. The nonadiabatic metric is related to but different from the quantum metric $g_{ij} = \text{Re} \mathcal{Q}_{ij}$ that is the real part of the quantum geometric tensor (or the Fubini-Study metric) 
\begin{align}
    \mathcal{Q}_{ij}= \sum_{m\neq 0} {A_{i,[0m]} A_{j,[m0]}}.
    \label{eq:quantumMetric}
\end{align}
In a two-band model with an energy gap $\Delta(\bm{q})$, $G_{ij}$ of the lower band is proportional to $g_{ij}$: $G_{ij} = \frac{2}{\Delta}g_{ij}$, which is positive-semidefinite. For the upper band, the nonadiabatic metric is negative-semidefinite. 
The energy denominator makes $G_{ij}$ not necessarily positive-semidefinite and thus not necessarily a Riemannian metric. 

Just as adiabatic wave packet theory unifies Berry phase effects through the term $\dot{q}^i A_i$~\cite{xiao2010berry}, the nonadiabatic corrections provide a comprehensive framework for capturing a wide array of phenomena that extend beyond the Berry phase effects. Below, we discuss the nonadiabatic effects in transport and quantum dynamics of Bloch electrons.

\section{Nonadiabatic Origin of Quantum-Metric-Related Transport} \label{sec:results}

\subsection{Nonadiabatic Velocity Corrections}\label{sec:resultsA}
Quantum-metric contributions have been identified in diverse electronic transport phenomena, ranging from nonlinear responses~\cite{yu2024quantum, liu2025quantum, verma2025quantum} to quantum capacitance~\cite{komissarov2024quantum}.
Yet, in contrast to Berry-phase physics, where wave packet dynamics supply a unified and intuitive interpretive framework, no analogous picture exists for understanding the origin of these quantum-metric-related effects. 
Our nonadiabatic theory provides such a framework. Specifically, the nonadiabatic metric contributes directly to the wave-packet velocity, as expressed in the following equations~\cite{SM}: 
\begin{align} \label{eq:velocity}
\dot{q}^i &= - \frac{\partial E}{\partial x_i} \\
\dot{x}_i &= \frac{\partial E}{\partial q^i} - F_{ij} \dot{q}^j + \Gamma_{i,jk} \dot{q}^j \dot{q}^k + G_{ij} \ddot{q}^j 
\end{align}
where the velocity $\dot{x}_i$ has four different contributions. The first term is the group velocity from band dispersion that was derived back to the 1930s~\cite{jones1934general, sundaram1999wave}. The second term is the anomalous velocity from the Berry curvature $F_{ij}=\partial_{q^i} {A}_j - \partial_{q^j} {A}_i$ arising from the adiabatic evolution of wave packets advanced by Chang, Sundaram, and Niu~\cite{chang1995berry, chang1996berry, sundaram1999wave}. 

The nonadiabatic metric term behaves like a kinetic energy of the generalized coordinate $\bm{q}$ in a curved space. It leads to two new velocity corrections.
The first velocity correction $\Gamma_{i,jk} \dot{q}^j \dot{q}^k$ is termed the {geodesic velocity}, as the coefficient $\Gamma_{i,jk}$ is closely related to $\Gamma^i_{jk}$ that defines the geodesic motion as shown later in Eq.~\eqref{eq:geodesicequation}.
$\Gamma_{i,jk}$ is the Christoffel symbol of the first kind:
\begin{align}
\Gamma_{i,jk} &= \frac{1}{2}  \left( \frac{\partial G_{ij}}{\partial q^k} + \frac{\partial G_{ik}}{\partial q^j} - \frac{\partial G_{jk}}{\partial q^i} \right).
\end{align}
The second nonadiabatic velocity $G_{ij} \ddot{q}^j $ is termed the geometric velocity, as the coefficient is the nonadiabatic metric $G_{ij}$. These velocity corrections arise solely from the time dependence of the parameter $\bm{q}$, and do not rely on the specific physical origin of the time variation of $\bm{q}$.

\subsection{Geodesic Velocity and Nonlinear Transport}
In the presence of a static electric field $\bm{\varepsilon}$, $\dot{\bm{q}}=\bm{\varepsilon}$ and $\ddot{\bm{q}}$ vanishes (i.e., we set the electron charge $e$ to unity). The velocity thus reduces to 
\begin{align} \label{eq:nonlineardotx}
    \dot{x}_i = \frac{\partial E}{\partial q^i} - F_{ij} \varepsilon^j + \Gamma_{i,jk} \varepsilon^j \varepsilon^k,
\end{align}
which has been widely adopted in studying nonlinear responses~\cite{gao2014field, liu2022berry, komissarov2024quantum, yu2024quantum, liu2025quantum, verma2025quantum}. 
By combining the velocity with the distribution function $f(\bm{q})$ obtained from the Boltzmann transport theory, one can calculate the current $j_i=\int \dot{x}_i(\bm{q}) f(\bm{q}) d\bm{q}$ induced by the electric field. 
Under the relaxation time approximation, the distribution function reads $f(\bm{q})=f_0(\bm{q}-\tau \dot{\bm{q}})$ where $f_0$ is the Fermi-Dirac distribution function and $\tau$ is the relaxation time. Since both the velocity and the distribution function can be expanded as a Taylor series of $\dot{\bm{q}}$, $f(\bm{q})=f_0(\bm{q})-\partial_{\bm{q}}f_0 \tau\dot{\bm{q}} + \cdots$ with $\dot{\bm{q}}=\bm{\varepsilon}$,
one can express the current to the linear and quadratic order of the field and thereby the linear and nonlinear responses.
It is noted that the velocity in Eq.~$\eqref{eq:nonlineardotx}$ was originally derived by studying adiabatic evolution $\cite{gao2014field, liu2022berry}$, using an asymptotic expansion in \textit{external} field strengths. In contrast, our derivation uses an expansion in the time derivatives of \textit{intrinsic} dynamical variables, which reproduces previous results in the static field limit. 
Although the coefficient in Eq.~\eqref{eq:nonlineardotx} was previously recognized as a Christoffel symbol and inspired studies of momentum space gravity~\cite{smith2022momentum, jain2024anomalous, mehraeen2025quantum}, its geometric origin remained obscure. The nonadiabatic metric introduced here provides a transparent geometric interpretation.

\subsection{Geometric Velocity and Electric Susceptibility}
The geometric velocity is proportional to $\ddot{\bm{q}}$, the acceleration of the generalized coordinate $\bm{q}$. While $\ddot{\bm{q}}$ vanishes under static fields or is negligible in the adiabatic limit, it becomes significant in nonadiabatic responses to time-dependent, oscillating forces. 
For instance, in the presence of a finite-frequency electric field $\varepsilon^i = \varepsilon_0^i e^{i\omega t}$ with $\varepsilon_0^i$ the amplitude and $\omega$ the frequency, the geometric velocity acquires a nonzero value: $G_{ij}\ddot{q}^j = G_{ij} \dot{\varepsilon}^j$ as $\dot{q}^i={\varepsilon}^i$. In an insulator, this velocity generates a current $j_i = \sum_n \int G_{ij,[n]}(\bm{q}) \dot{\varepsilon}^j d\bm{q} = \partial_t (\chi_{ij}\varepsilon^j)$ where $G_{ij,[n]}$ is the nonadiabatic metric for the $n$-th occupied band. This current corresponds to the dissipationless polarization current where $\chi_{ij}=\sum_n \int G_{ij,[n]} d\bm{q}$ is the electric susceptibility~\cite{komissarov2024quantum, resta2025nonadiabatic}. The results agree with the linear response theory based on the Kubo formula~\cite{komissarov2024quantum, resta2025nonadiabatic} and more details for a metallic system are discussed in Sec. III in the Supplemental Material~\cite{SM}.

Thus, the integral of the nonadiabatic metric over the Brillouin zone is related to the electric susceptibility. This is in contrast to the quantum metric, the integral of which over the Brillouin zone is related to the spread of the Wannier function~\cite{marzari1997maximally}. The full implication of the geometric velocity remains an open area for further investigation. It will be manifested in nonlinear and nonadiabatic responses to finite-frequency driving, such as below-gap light~\cite{oka2019floquet,zhou2023floquet,zhan2024perspective,choi2025observation} and coherently excited collective excitations such as magnons and phonons~\cite{nakata2017magnonic, li2021topological, tang2024lossless, cheng2020large, basini2024terahertz, xu2025time}. 
Those processes are broadly relevant for phenomena such as second-harmonic generation~\cite{bhalla2022resonant} and nonlinear phononics~\cite{nakata2017magnonic, li2021topological, tang2024lossless, cheng2020large, basini2024terahertz, xu2025time}. 

This unified framework emphasizes the fundamental role of the nonadiabatic metric $G_{ij}$. While earlier work associated $G_{ij}$ with Berry connection polarizability induced by external fields~\cite{gao2014field, liu2022berry, komissarov2024quantum}, our results establish its inherently nonadiabatic nature, emerging from the interband processes due to the dynamical changing of intrinsic parameters.

\section{Momentum-Space Metric Tensor}\label{sec:qspace}

\subsection{Metric Tensor and Forced Geodesic Equation}\label{sec:resultsC} 
The nonadiabatic metric also underscores the importance of the $\bm{q}$-space picture, instead of the real space, where $\bm{q}$ is taken as a generalized coordinate to describe the wave packet dynamics. This enables us to study the emergent gravity and emergent mass in momentum space, which will be discussed in this and following subsections.

$G_{ij}(\bm{q})$ in the Lagrangian defines a metric tensor in the tangent vector space of the generalized velocity $\dot{\bm{q}}$ defining a line element \( ds^2 = G_{ij} dq^i dq^j \). It defines the geometric distance between the momentum-space centers of two wave packets.
This is in contrast to the quantum metric that defines the quantum distance $ds_Q$ between two wave functions $|u_{\bm{q}}\rangle$ and $|u_{\bm{q}+d\bm{q}}\rangle$ at $\bm{q}$ and $\bm{q}+d\bm{q}$ via $ds_Q^2=1-|\langle u_{\bm{q}} | u_{\bm{q}+d\bm{q}} \rangle|^2$~\cite{provost1980riemannian}.
A \(\bm{q}\)-dependent \( G_{ij} \) signifies a curved manifold.
Conversely, in cases where \( G_{ij} \) is constant across the momentum space, it corresponds to an effective mass in a flat \(\bm{q}\)-space, simplifying the dynamics to a Euclidean framework. 

\begin{figure}[t!]
\includegraphics[width=1\linewidth]{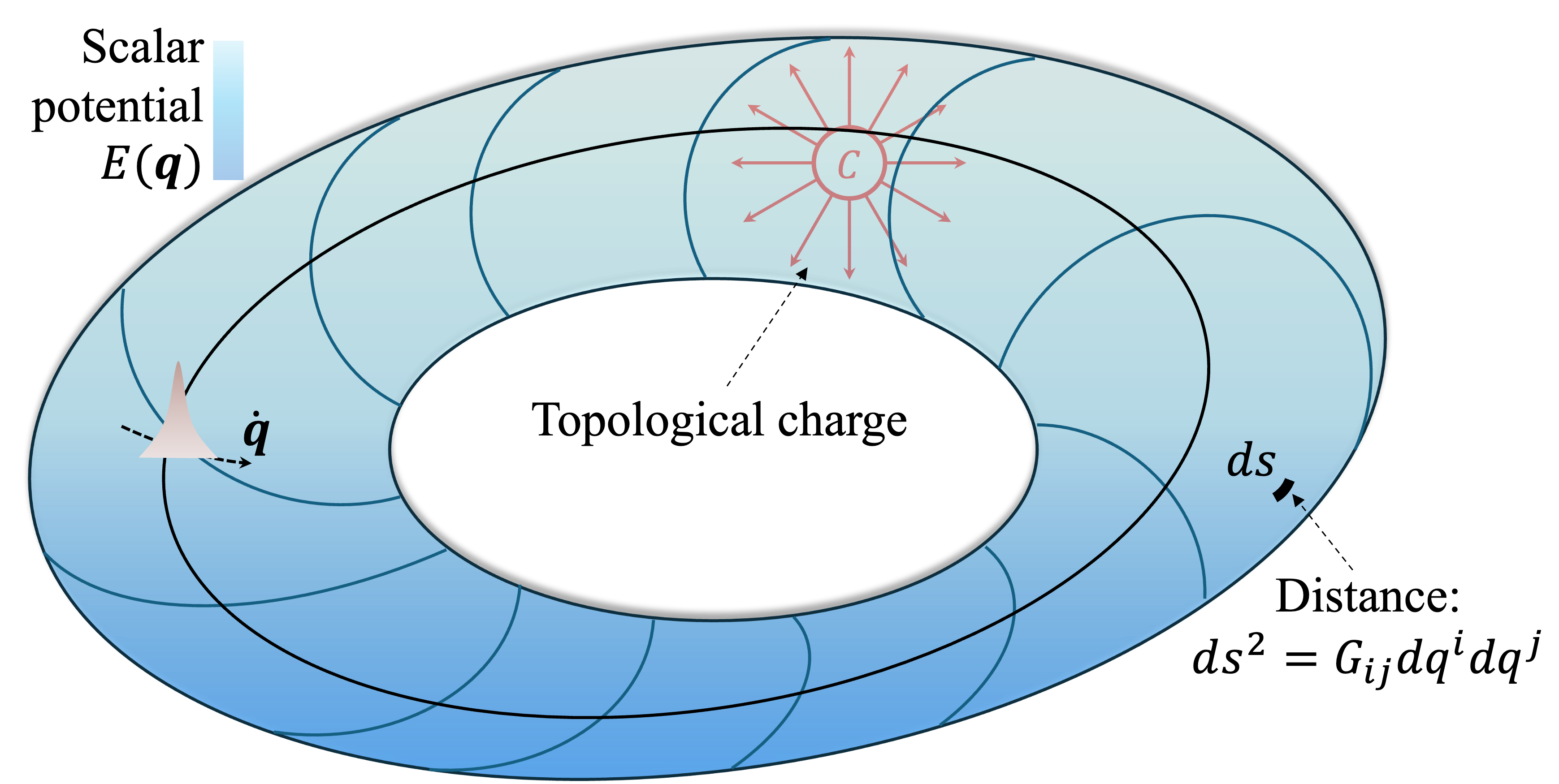}
\caption{Wave packet dynamics in $\bm{q}$-space, a torus with curved geometry. The distance $ds^2$ is defined by the nonadiabatic metric $G_{ij}$. The topological charge stands for the Dirac monopole that generates nonzero gauge fields corresponding to the Berry curvature. $C$ indicates the Chern number. Color encodes the scalar potential from the band dispersion.}
\label{fig:geodesic}
\end{figure}

Expressing the equations of motion in the $\bm{q}$-space clarifies the geometric interpretation. When the inverse of the metric tensor exists, the equation of motion reads
\begin{align} \label{eq:geodesicequation}
\ddot{q}^i + \Gamma^i_{jk} \dot{q}^j \dot{q}^k = & G^{ij}\left(-\frac{\partial E}{\partial q^j} + F_{jk} \dot{q}^k + \dot{x}_j \right) 
\end{align}
where $\Gamma^l_{jk}$ is the Christoffel symbol of the second kind
\begin{align}\label{eq:christoffel2nd}
    \Gamma^i_{jk}=G^{il}\Gamma_{l,jk}=\frac{1}{2} G^{i l} \left( \frac{\partial G_{lj}}{\partial q^k} + \frac{\partial G_{kl}}{\partial q^j} - \frac{\partial G_{jk}}{\partial q^l}  \right),
\end{align}
$G^{ji}G_{ik}=\delta^j_k$ with $\delta^j_k$ the Kronecker delta.
When the right-hand side of Eq.~\eqref{eq:geodesicequation} vanishes, the equation reduces to the geodesic equation for a free particle navigating a curved manifold with metric tensor $G_{ij}$. A nonzero right-hand side indicates a particle subject to forces from three sources (see Fig.~\ref{fig:geodesic}): a scalar potential from the band dispersion $E$, a gauge field from the Berry curvature $F_{ij}$, and an external force from the wave packet velocity $\dot{x}_j$. This equation enables the study of gravity in condensed matter setups~\cite{volovik2003universe}. It is noted that Eq.~\eqref{eq:geodesicequation} differs from the recent study on momentum space gravity~\cite{smith2022momentum, jain2024anomalous, mehraeen2025quantum} where $G^{ij}$ and $\ddot{\bm{q}}$ are absent and $\Gamma_{i,jk}$ is used instead of $\Gamma^i_{jk}$.

When $x_i$ can be invertibly represented as a function of $\dot{q}^j$ by using $\dot{q}^i = - \frac{\partial E}{\partial x_i}$, we have 
\begin{align}
    x_i = x_i(\dot{\bm{q}}), ~~~
    \dot{x}_i = \frac{\partial x_i}{\partial \dot{q}^j} \ddot{q}^j =- \Xi_{ij} \ddot{q}^j,
\end{align}
the equations of motion find their $\bm{q}$-space representation:
\begin{align} \label{eq:geodesicequation1}
\ddot{q}^i + \bar{\Gamma}^i_{jk} \dot{q}^j \dot{q}^k = & \bar{G}^{ij}\left(-\frac{\partial E}{\partial q^j} + F_{jk} \dot{q}^k \right),
\end{align}
where $\bar{G}_{ij}=G_{ij} + \Xi_{ij}$, $\bar{G}^{ij}\bar{G}_{jk}=\delta^i_k$, and $\bar{\Gamma}^i_{jk}$ is the corresponding Christoffel symbol. With information of $\dot{\bm{q}}$, $x_i$ can be solved using algebraic equations.
The metric tensor becomes dynamical with 
\begin{align} \label{eq:metricdynamics}
    \dot{\bar{G}}_{ij}=\frac{\partial G_{ij}}{\partial q_k}\dot{q}_k + \frac{\partial \Xi_{ij}}{\partial \dot{q}_k}\ddot{q}_k.
\end{align}
Unlike general relativity, the effective metric here is not an independent dynamical field but an induced object determined by the underlying microscopic degrees of freedom.

\subsection{Massive Particle Dynamics in Flat Bands}\label{sec:resultsB}
The re-quantization of semiclassical wave packet dynamics is fruitful in describing various quantum phenomena such as Landau levels of Bloch electrons~\cite{chang2008berry} and Berry curvature effects on exciton spectrum~\cite{zhou2015berry} and Wigner crystal phases~\cite{shi2018dynamics}. While the importance of adiabatic Berry phase on those phenomena has attracted much attention in emerging material systems, such as transition-metal dichalcogenides, gapped graphenes, Landau levels, and moir\'e systems~\cite{zhou2015berry, srivastava2015signatures, skinner2016interlayer, wu2017topological, shi2018dynamics, joy2025chiral, price2014quantum, dong2025phonons}, the effects of nonadiabatic corrections are yet to be explored.

We focus on the lowest Landau level as a concrete example. The simplicity enables analytical insights that can be generalized qualitatively to other flat-band systems. For the lowest Landau level, the Berry curvature $F_{ij}$ and nonadiabatic metric $G_{ij}$ are constants. Consider a rectangular Brillouin zone with dimensions $L_{q_{x}, q_y} = 2\pi/L_{x,y}$ where $L_{q_{x}, q_y}$ defines the ${\bm{q}}$-space size while $L_{x,y}$ are the dimensions of real-space unit cell with one magnetic flux quantum. The Berry curvature is uniform and its integral over the Brillouin zone corresponds to the Chern number $C$, i.e., $F =\frac{2\pi C}{L_{q_x}L_{q_y}}$. 
The nonadiabatic metric reads $G_{ij}=G_0 \mathcal{I}_{ij}$ with $\mathcal{I}$ the identity matrix and $G_{0}=\frac{|F|}{\Delta}$ where $\Delta$ is the energy spacing between the lowest and the first Landau levels (see Supplemental Material)~\cite{SM}. 

We consider the wave packet dynamics in a harmonic potential that can arise from external electrostatic fields or Coulomb interactions in Wigner crystals~\cite{shi2018dynamics, joy2025chiral}
\begin{align} \label{eq:HarmonicPotential}
    V(\bm{x})=\frac{\kappa}{2} \bm{x}^2.
\end{align}
Thus, we have the equation of motion $\dot{q}^i = - \kappa x^i $ and thereby $\dot{x}^i = -\frac{1}{\kappa}\ddot{q}^i$. Here $x^i = \delta^{ij} x_j$ and $\delta^{ij}$ the Kronecker delta. The equation of $\bm{q}$ can be expressed as a forced geodesic equation in the momentum space ($\bm{q}$-space):
\begin{align} \label{eq:geodesicequation0}
\ddot{q}^i + \bar{\Gamma}^i_{jk} \dot{q}^j \dot{q}^k = & \bar{G}^{ij}\left(-\frac{\partial E}{\partial q^j} + F_{jk} \dot{q}^k \right),
\end{align}
where $\bar{G}_{ij}=G_{ij} + \frac{1}{\kappa}$, $\bar{G}^{ij}\bar{G}_{jk}=\delta^i_k$, and $\bar{\Gamma}^i_{jk}$ is the corresponding Christoffel symbol that vanishes in this example as $G_{ij}$ is constant. The group velocity vanishes as $E$ is a constant and $F_{jk}$ corresponds to the artificial magnetic field in the momentum space from Berry curvature.
This equation follows the Lagrangian~\cite{SM} 
\begin{equation}
L = \frac{1}{2} \bar{G}_{ij} \dot{q}^i \dot{q}^j -A_i \dot{q}^i 
\end{equation}
where \( A_j \) is a vector potential and the band energy is a constant that is set as zero.
This Lagrangian has a well-defined canonical structure enabling us to perform canonical quantization. The canonical momentum is $p_i = \frac{\partial L}{\partial \dot{q}^i} =\bar{G}_{ij} \dot{q}^j - A_i$
and the Hamiltonian reads
\begin{equation} \label{eq:Ham0}
H = \frac{1}{2M} \delta^{ij}(p_i + A_i)(p_j + A_j) 
\end{equation}
describing the dynamics of a massive particle under electromagnetic fields on a flat torus with $M=G_0 + \frac{1}{\kappa}$. By promoting $(p_i, q^j)$ to operators with the commutator $[\hat{p}_i, \hat{q}^j]=-i\delta_i^j$, we obtain the quantized Hamiltonian in momentum-space representation. We note that here the canonical momentum $\hat{p}_i$ corresponds to the real-space coordinate, ${\bm{x}} = -\frac{1}{\kappa}\dot{\bm{q}} = -\frac{1}{1+\kappa G_0}({\bm{p}}+{\bm{A}})$, consistent with the canonical variables in Ref.~\onlinecite{zhou2015berry}, but with the distinction that it incorporates a nonadiabatic correction from $\kappa G_0$.

The topological structure of the Brillouin zone as a torus requires that the $\bm{q}$-space wave functions satisfy periodic boundary conditions. 
This boundary condition is compatible with the quantization of the Chern number, enabling a direct connection between the electronic states in $\bm{q}$-space to the quantum Hall effects on a torus~\cite{onofri2001landau, fremling2015quantum}. 
The eigenenergies are $\epsilon_n=(n+\frac{1}{2})\frac{F}{M}=(n+\frac{1}{2})\frac{F}{1/\kappa+G_0}$ with each level's degeneracy equal to $|C|$. The confining potential $\kappa$ corresponds to an effective inverse mass that is modified by the nonadiabatic metric $G_0$ from interband mixing. The effect of $G_0$ is enhanced by increasing $\kappa$, i.e., increasing the potential strength, which can increase $\dot{\bm{q}}$ and thus the interband contributions.

\begin{figure}[t!]
\includegraphics[width=1\linewidth]{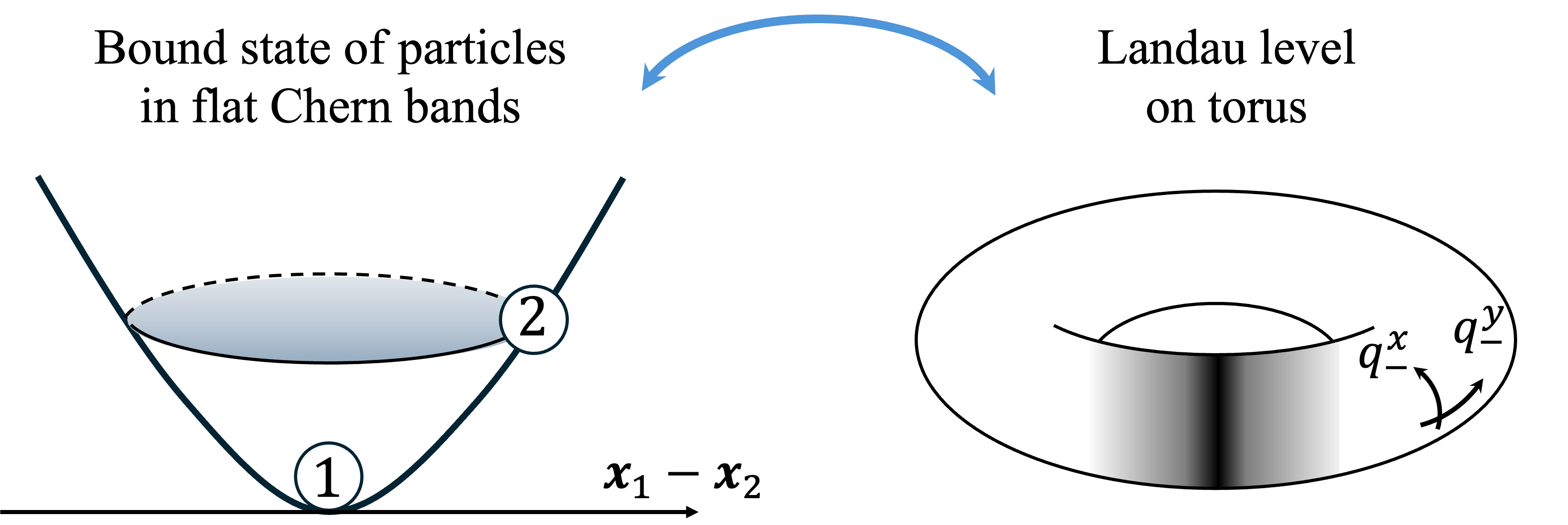}
\caption{Bound states of flat-band particles in real space mirror the Landau-level wave functions on a torus in $\bm{q}$-space.}
\label{fig:cooper}
\end{figure}

\subsection{Two-Body Bound States in Flat Bands}
The confining problem can also be generalized to two-body bound states~\cite{zhou2015berry} where the effect of $G_0$ can be enhanced as $\kappa$ originates from the Coulomb interaction strength. Here we consider two wave packets in flat bands. We assume that the two particles live in two layers well separated by a distance $x_0$, and they interact through a Coulomb interaction in real space.
The two-particle Lagrangian is
\begin{align}
L_{\rm T}=L({\bm{x}}_1) + L({\bm{x}}_2) + \frac{1}{\epsilon \sqrt{(\bm{x}_1 - \bm{x}_2)^2+x_0^2}}
\end{align}
where $\bm{x}_{1,2}$ are wave packet centers, $L(\bm{x}_{1,2})$ is the single-particle Lagrangian with the same constant Berry curvature $F$ and nonadiabatic metric $G_0$, $\epsilon$ is the effective dielectric constant, the charge is unity, and $x_0$ is the cutoff length.
Focusing on bound states with $|\bm{x}_1 - \bm{x}_2| \ll x_0$, we approximate the Lagrangian as
\begin{align}
L_{\rm T}\simeq L({\bm{x}}_1) + L({\bm{x}}_2) + \left[ V_0 - \frac{1}{2}\kappa_{\rm I} ({\bm{x}}_1 - {\bm{x}}_2)^2 \right] 
\end{align}
where $V_0 = \frac{1}{\epsilon x_0}$ and $\kappa_{\rm I}=\frac{V_0}{x_0^2}$.

We focus on the relative dynamics. Applying canonical quantization, the effective Hamiltonian for the relative coordinate is
\begin{equation} \label{eq:cooperpair}
\hat{H}_{\rm -} = \frac{1}{2M'}(\hat{\bm{p}}_- +\hat{\bm{A}}(\hat{\bm{q}}_-))^2 - V_0
\end{equation}
where $\hat{\bm{q}}_- = (\hat{\bm{q}}_1 - \hat{\bm{q}}_2)/\sqrt{2}$, $[\hat{p}_{-,i}, \hat{q}_-^j]=-i\delta_i^j$, and $M'=G_0+ \frac{1}{2\kappa_{\rm I}}$ is the reduced effective mass. This Hamiltonian again resembles a Landau level problem on a torus (see Fig. 3) with eigenenergies $\epsilon_n=(n+\frac{1}{2})\frac{F}{M'}- V_0$.

If the two particles are different, like an electron and a hole, the wave function of the relative coordinate is only restricted by the periodic boundary condition. 
Here we consider ideal situations to illustrate the effects of the nonadiabatic metric. Their effects in more realistic conditions deserve detailed study in the future, which are naturally related to excitons formed by electrons and holes in flat bands~\cite{eisenstein2014exciton, zhou2015berry, verma2024geometric}. The potential relationship with Cooper pairs is interesting and warrants further investigation~\cite{torma2018quantum, cao2018unconventional, han2024signatures, xu2025signatures}.

\section{Summary} \label{sec:summary}
We introduce a geometric framework that extends the study of Bloch electron dynamics in a scalar potential beyond adiabatic Berry-phase effects. We show that the leading-order nonadiabatic effects are elegantly characterized by the nonadiabatic metric---a momentum-space metric tensor that is related to but distinct from the quantum metric. This nonadiabatic metric unifies the nonlinear and nonadiabatic responses by introducing geometric and geodesic velocities beyond the anomalous velocity from the Berry curvature. The geometric velocity is related to the metric tensor, whereas the geodesic velocity is related to the Christoffel symbol. The nonadiabatic metric also modifies the quantum dynamics of electrons, manifested as a metric in momentum space. A constant nonadiabatic metric influences the flat-band electron dynamics in a harmonic potential by introducing a correction to the effective mass for the generalized coordinate $\bm{q}$. 
In general, the nonadiabatic metric is not a constant, which endows momentum space with a curved geometry, making the electronic dynamics a forced geodesic motion in momentum space. 

\section*{Acknowledgments} 
Y.R. thanks Qian Niu, Yang Gao, Di Xiao, Wenqin Chen and Jie Wang for useful discussions.


\begin{thebibliography}{92}%
\makeatletter
\providecommand \@ifxundefined [1]{%
 \@ifx{#1\undefined}
}%
\providecommand \@ifnum [1]{%
 \ifnum #1\expandafter \@firstoftwo
 \else \expandafter \@secondoftwo
 \fi
}%
\providecommand \@ifx [1]{%
 \ifx #1\expandafter \@firstoftwo
 \else \expandafter \@secondoftwo
 \fi
}%
\providecommand \natexlab [1]{#1}%
\providecommand \enquote  [1]{``#1''}%
\providecommand \bibnamefont  [1]{#1}%
\providecommand \bibfnamefont [1]{#1}%
\providecommand \citenamefont [1]{#1}%
\providecommand \href@noop [0]{\@secondoftwo}%
\providecommand \href [0]{\begingroup \@sanitize@url \@href}%
\providecommand \@href[1]{\@@startlink{#1}\@@href}%
\providecommand \@@href[1]{\endgroup#1\@@endlink}%
\providecommand \@sanitize@url [0]{\catcode `\\12\catcode `\$12\catcode `\&12\catcode `\#12\catcode `\^12\catcode `\_12\catcode `\%12\relax}%
\providecommand \@@startlink[1]{}%
\providecommand \@@endlink[0]{}%
\providecommand \url  [0]{\begingroup\@sanitize@url \@url }%
\providecommand \@url [1]{\endgroup\@href {#1}{\urlprefix }}%
\providecommand \urlprefix  [0]{URL }%
\providecommand \Eprint [0]{\href }%
\providecommand \doibase [0]{https://doi.org/}%
\providecommand \selectlanguage [0]{\@gobble}%
\providecommand \bibinfo  [0]{\@secondoftwo}%
\providecommand \bibfield  [0]{\@secondoftwo}%
\providecommand \translation [1]{[#1]}%
\providecommand \BibitemOpen [0]{}%
\providecommand \bibitemStop [0]{}%
\providecommand \bibitemNoStop [0]{.\EOS\space}%
\providecommand \EOS [0]{\spacefactor3000\relax}%
\providecommand \BibitemShut  [1]{\csname bibitem#1\endcsname}%
\let\auto@bib@innerbib\@empty
\bibitem [{\citenamefont {{Berry}}(1984)}]{berry1984quantal}%
  \BibitemOpen
  \bibfield  {author} {\bibinfo {author} {\bibfnamefont {M.~V.}\ \bibnamefont {{Berry}}},\ }\bibfield  {title} {\bibinfo {title} {Quantal phase factors accompanying adiabatic changes},\ }\href@noop {} {\bibfield  {journal} {\bibinfo  {journal} {Proceedings of the Royal Society of London. A. Mathematical and Physical Sciences}\ }\textbf {\bibinfo {volume} {392}},\ \bibinfo {pages} {45} (\bibinfo {year} {1984})}\BibitemShut {NoStop}%
\bibitem [{\citenamefont {Xiao}\ \emph {et~al.}(2010)\citenamefont {Xiao}, \citenamefont {Chang},\ and\ \citenamefont {Niu}}]{xiao2010berry}%
  \BibitemOpen
  \bibfield  {author} {\bibinfo {author} {\bibfnamefont {D.}~\bibnamefont {Xiao}}, \bibinfo {author} {\bibfnamefont {M.-C.}\ \bibnamefont {Chang}},\ and\ \bibinfo {author} {\bibfnamefont {Q.}~\bibnamefont {Niu}},\ }\bibfield  {title} {\bibinfo {title} {{Berry} phase effects on electronic properties},\ }\href@noop {} {\bibfield  {journal} {\bibinfo  {journal} {Rev. Mod. Phys.}\ }\textbf {\bibinfo {volume} {82}},\ \bibinfo {pages} {1959} (\bibinfo {year} {2010})}\BibitemShut {NoStop}%
\bibitem [{\citenamefont {Provost}\ and\ \citenamefont {Vallee}(1980)}]{provost1980riemannian}%
  \BibitemOpen
  \bibfield  {author} {\bibinfo {author} {\bibfnamefont {J.}~\bibnamefont {Provost}}\ and\ \bibinfo {author} {\bibfnamefont {G.}~\bibnamefont {Vallee}},\ }\bibfield  {title} {\bibinfo {title} {Riemannian structure on manifolds of quantum states},\ }\href@noop {} {\bibfield  {journal} {\bibinfo  {journal} {Communications in Mathematical Physics}\ }\textbf {\bibinfo {volume} {76}},\ \bibinfo {pages} {289} (\bibinfo {year} {1980})}\BibitemShut {NoStop}%
\bibitem [{\citenamefont {Parameswaran}\ \emph {et~al.}(2013)\citenamefont {Parameswaran}, \citenamefont {Roy},\ and\ \citenamefont {Sondhi}}]{parameswaran2013fractional}%
  \BibitemOpen
  \bibfield  {author} {\bibinfo {author} {\bibfnamefont {S.~A.}\ \bibnamefont {Parameswaran}}, \bibinfo {author} {\bibfnamefont {R.}~\bibnamefont {Roy}},\ and\ \bibinfo {author} {\bibfnamefont {S.~L.}\ \bibnamefont {Sondhi}},\ }\bibfield  {title} {\bibinfo {title} {Fractional quantum hall physics in topological flat bands},\ }\href@noop {} {\bibfield  {journal} {\bibinfo  {journal} {Comptes Rendus Physique}\ }\textbf {\bibinfo {volume} {14}},\ \bibinfo {pages} {816} (\bibinfo {year} {2013})}\BibitemShut {NoStop}%
\bibitem [{\citenamefont {Peotta}\ and\ \citenamefont {T{\"o}rm{\"a}}(2015)}]{peotta2015superfluidity}%
  \BibitemOpen
  \bibfield  {author} {\bibinfo {author} {\bibfnamefont {S.}~\bibnamefont {Peotta}}\ and\ \bibinfo {author} {\bibfnamefont {P.}~\bibnamefont {T{\"o}rm{\"a}}},\ }\bibfield  {title} {\bibinfo {title} {Superfluidity in topologically nontrivial flat bands},\ }\href@noop {} {\bibfield  {journal} {\bibinfo  {journal} {Nature communications}\ }\textbf {\bibinfo {volume} {6}},\ \bibinfo {pages} {8944} (\bibinfo {year} {2015})}\BibitemShut {NoStop}%
\bibitem [{\citenamefont {T{\"o}rm{\"a}}(2023)}]{torma2023essay}%
  \BibitemOpen
  \bibfield  {author} {\bibinfo {author} {\bibfnamefont {P.}~\bibnamefont {T{\"o}rm{\"a}}},\ }\bibfield  {title} {\bibinfo {title} {Essay: Where can quantum geometry lead us?},\ }\href@noop {} {\bibfield  {journal} {\bibinfo  {journal} {Physical Review Letters}\ }\textbf {\bibinfo {volume} {131}},\ \bibinfo {pages} {240001} (\bibinfo {year} {2023})}\BibitemShut {NoStop}%
\bibitem [{\citenamefont {Marzari}\ and\ \citenamefont {Vanderbilt}(1997)}]{marzari1997maximally}%
  \BibitemOpen
  \bibfield  {author} {\bibinfo {author} {\bibfnamefont {N.}~\bibnamefont {Marzari}}\ and\ \bibinfo {author} {\bibfnamefont {D.}~\bibnamefont {Vanderbilt}},\ }\bibfield  {title} {\bibinfo {title} {Maximally localized generalized {Wannier} functions for composite energy bands},\ }\href@noop {} {\bibfield  {journal} {\bibinfo  {journal} {Phys. Rev. B}\ }\textbf {\bibinfo {volume} {56}},\ \bibinfo {pages} {12847} (\bibinfo {year} {1997})}\BibitemShut {NoStop}%
\bibitem [{\citenamefont {Yu}\ \emph {et~al.}(2024)\citenamefont {Yu}, \citenamefont {Bernevig}, \citenamefont {Queiroz}, \citenamefont {Rossi}, \citenamefont {T{\"o}rm{\"a}},\ and\ \citenamefont {Yang}}]{yu2024quantum}%
  \BibitemOpen
  \bibfield  {author} {\bibinfo {author} {\bibfnamefont {J.}~\bibnamefont {Yu}}, \bibinfo {author} {\bibfnamefont {B.~A.}\ \bibnamefont {Bernevig}}, \bibinfo {author} {\bibfnamefont {R.}~\bibnamefont {Queiroz}}, \bibinfo {author} {\bibfnamefont {E.}~\bibnamefont {Rossi}}, \bibinfo {author} {\bibfnamefont {P.}~\bibnamefont {T{\"o}rm{\"a}}},\ and\ \bibinfo {author} {\bibfnamefont {B.-J.}\ \bibnamefont {Yang}},\ }\bibfield  {title} {\bibinfo {title} {Quantum geometry in quantum materials},\ }\href@noop {} {\bibfield  {journal} {\bibinfo  {journal} {arXiv preprint arXiv:2501.00098}\ } (\bibinfo {year} {2024})}\BibitemShut {NoStop}%
\bibitem [{\citenamefont {Liu}\ \emph {et~al.}(2025)\citenamefont {Liu}, \citenamefont {Qiang}, \citenamefont {Lu},\ and\ \citenamefont {Xie}}]{liu2025quantum}%
  \BibitemOpen
  \bibfield  {author} {\bibinfo {author} {\bibfnamefont {T.}~\bibnamefont {Liu}}, \bibinfo {author} {\bibfnamefont {X.-B.}\ \bibnamefont {Qiang}}, \bibinfo {author} {\bibfnamefont {H.-Z.}\ \bibnamefont {Lu}},\ and\ \bibinfo {author} {\bibfnamefont {X.}~\bibnamefont {Xie}},\ }\bibfield  {title} {\bibinfo {title} {Quantum geometry in condensed matter},\ }\href@noop {} {\bibfield  {journal} {\bibinfo  {journal} {National Science Review}\ }\textbf {\bibinfo {volume} {12}},\ \bibinfo {pages} {nwae334} (\bibinfo {year} {2025})}\BibitemShut {NoStop}%
\bibitem [{\citenamefont {Verma}\ \emph {et~al.}(2025)\citenamefont {Verma}, \citenamefont {Moll}, \citenamefont {Holder},\ and\ \citenamefont {Queiroz}}]{verma2025quantum}%
  \BibitemOpen
  \bibfield  {author} {\bibinfo {author} {\bibfnamefont {N.}~\bibnamefont {Verma}}, \bibinfo {author} {\bibfnamefont {P.~J.}\ \bibnamefont {Moll}}, \bibinfo {author} {\bibfnamefont {T.}~\bibnamefont {Holder}},\ and\ \bibinfo {author} {\bibfnamefont {R.}~\bibnamefont {Queiroz}},\ }\bibfield  {title} {\bibinfo {title} {Quantum geometry: Revisiting electronic scales in quantum matter},\ }\href@noop {} {\bibfield  {journal} {\bibinfo  {journal} {arXiv preprint arXiv:2504.07173}\ } (\bibinfo {year} {2025})}\BibitemShut {NoStop}%
\bibitem [{\citenamefont {Neupert}\ \emph {et~al.}(2013)\citenamefont {Neupert}, \citenamefont {Chamon},\ and\ \citenamefont {Mudry}}]{neupert2013measuring}%
  \BibitemOpen
  \bibfield  {author} {\bibinfo {author} {\bibfnamefont {T.}~\bibnamefont {Neupert}}, \bibinfo {author} {\bibfnamefont {C.}~\bibnamefont {Chamon}},\ and\ \bibinfo {author} {\bibfnamefont {C.}~\bibnamefont {Mudry}},\ }\bibfield  {title} {\bibinfo {title} {Measuring the quantum geometry of {Bloch} bands with current noise},\ }\href@noop {} {\bibfield  {journal} {\bibinfo  {journal} {Phys. Rev. B}\ }\textbf {\bibinfo {volume} {87}},\ \bibinfo {pages} {245103} (\bibinfo {year} {2013})}\BibitemShut {NoStop}%
\bibitem [{\citenamefont {Gao}\ \emph {et~al.}(2014)\citenamefont {Gao}, \citenamefont {Yang},\ and\ \citenamefont {Niu}}]{gao2014field}%
  \BibitemOpen
  \bibfield  {author} {\bibinfo {author} {\bibfnamefont {Y.}~\bibnamefont {Gao}}, \bibinfo {author} {\bibfnamefont {S.~A.}\ \bibnamefont {Yang}},\ and\ \bibinfo {author} {\bibfnamefont {Q.}~\bibnamefont {Niu}},\ }\bibfield  {title} {\bibinfo {title} {Field induced positional shift of {Bloch} electrons and its dynamical implications},\ }\href@noop {} {\bibfield  {journal} {\bibinfo  {journal} {Phys. Rev. Lett.}\ }\textbf {\bibinfo {volume} {112}},\ \bibinfo {pages} {166601} (\bibinfo {year} {2014})}\BibitemShut {NoStop}%
\bibitem [{\citenamefont {Morimoto}\ and\ \citenamefont {Nagaosa}(2016)}]{morimoto2016topological}%
  \BibitemOpen
  \bibfield  {author} {\bibinfo {author} {\bibfnamefont {T.}~\bibnamefont {Morimoto}}\ and\ \bibinfo {author} {\bibfnamefont {N.}~\bibnamefont {Nagaosa}},\ }\bibfield  {title} {\bibinfo {title} {Topological nature of nonlinear optical effects in solids},\ }\href@noop {} {\bibfield  {journal} {\bibinfo  {journal} {Sci. Adv.}\ }\textbf {\bibinfo {volume} {2}},\ \bibinfo {pages} {e1501524} (\bibinfo {year} {2016})}\BibitemShut {NoStop}%
\bibitem [{\citenamefont {Ozawa}(2018)}]{ozawa2018steady}%
  \BibitemOpen
  \bibfield  {author} {\bibinfo {author} {\bibfnamefont {T.}~\bibnamefont {Ozawa}},\ }\bibfield  {title} {\bibinfo {title} {Steady-state {Hall} response and quantum geometry of driven-dissipative lattices},\ }\href@noop {} {\bibfield  {journal} {\bibinfo  {journal} {Phys. Rev. B}\ }\textbf {\bibinfo {volume} {97}},\ \bibinfo {pages} {041108} (\bibinfo {year} {2018})}\BibitemShut {NoStop}%
\bibitem [{\citenamefont {Gao}\ and\ \citenamefont {Xiao}(2019)}]{gao2019nonreciprocal}%
  \BibitemOpen
  \bibfield  {author} {\bibinfo {author} {\bibfnamefont {Y.}~\bibnamefont {Gao}}\ and\ \bibinfo {author} {\bibfnamefont {D.}~\bibnamefont {Xiao}},\ }\bibfield  {title} {\bibinfo {title} {Nonreciprocal directional dichroism induced by the quantum metric dipole},\ }\href@noop {} {\bibfield  {journal} {\bibinfo  {journal} {Phys. Rev. Lett.}\ }\textbf {\bibinfo {volume} {122}},\ \bibinfo {pages} {227402} (\bibinfo {year} {2019})}\BibitemShut {NoStop}%
\bibitem [{\citenamefont {Ahn}\ \emph {et~al.}(2020)\citenamefont {Ahn}, \citenamefont {Guo},\ and\ \citenamefont {Nagaosa}}]{ahn2020low}%
  \BibitemOpen
  \bibfield  {author} {\bibinfo {author} {\bibfnamefont {J.}~\bibnamefont {Ahn}}, \bibinfo {author} {\bibfnamefont {G.-Y.}\ \bibnamefont {Guo}},\ and\ \bibinfo {author} {\bibfnamefont {N.}~\bibnamefont {Nagaosa}},\ }\bibfield  {title} {\bibinfo {title} {Low-frequency divergence and quantum geometry of the bulk photovoltaic effect in topological semimetals},\ }\href@noop {} {\bibfield  {journal} {\bibinfo  {journal} {Phys. Rev. X}\ }\textbf {\bibinfo {volume} {10}},\ \bibinfo {pages} {041041} (\bibinfo {year} {2020})}\BibitemShut {NoStop}%
\bibitem [{\citenamefont {Ahn}\ \emph {et~al.}(2022)\citenamefont {Ahn}, \citenamefont {Guo}, \citenamefont {Nagaosa},\ and\ \citenamefont {Vishwanath}}]{ahn2022riemannian}%
  \BibitemOpen
  \bibfield  {author} {\bibinfo {author} {\bibfnamefont {J.}~\bibnamefont {Ahn}}, \bibinfo {author} {\bibfnamefont {G.-Y.}\ \bibnamefont {Guo}}, \bibinfo {author} {\bibfnamefont {N.}~\bibnamefont {Nagaosa}},\ and\ \bibinfo {author} {\bibfnamefont {A.}~\bibnamefont {Vishwanath}},\ }\bibfield  {title} {\bibinfo {title} {Riemannian geometry of resonant optical responses},\ }\href@noop {} {\bibfield  {journal} {\bibinfo  {journal} {Nat. Phys.}\ }\textbf {\bibinfo {volume} {18}},\ \bibinfo {pages} {290} (\bibinfo {year} {2022})}\BibitemShut {NoStop}%
\bibitem [{\citenamefont {Nag}\ \emph {et~al.}(2023)\citenamefont {Nag}, \citenamefont {Das}, \citenamefont {Zeng},\ and\ \citenamefont {Nandy}}]{nag2023third}%
  \BibitemOpen
  \bibfield  {author} {\bibinfo {author} {\bibfnamefont {T.}~\bibnamefont {Nag}}, \bibinfo {author} {\bibfnamefont {S.~K.}\ \bibnamefont {Das}}, \bibinfo {author} {\bibfnamefont {C.}~\bibnamefont {Zeng}},\ and\ \bibinfo {author} {\bibfnamefont {S.}~\bibnamefont {Nandy}},\ }\bibfield  {title} {\bibinfo {title} {Third-order {Hall} effect in the surface states of a topological insulator},\ }\href@noop {} {\bibfield  {journal} {\bibinfo  {journal} {Phys. Rev. B}\ }\textbf {\bibinfo {volume} {107}},\ \bibinfo {pages} {245141} (\bibinfo {year} {2023})}\BibitemShut {NoStop}%
\bibitem [{\citenamefont {Ma}\ \emph {et~al.}(2023)\citenamefont {Ma}, \citenamefont {Arora}, \citenamefont {Vignale},\ and\ \citenamefont {Song}}]{ma2023anomalous}%
  \BibitemOpen
  \bibfield  {author} {\bibinfo {author} {\bibfnamefont {D.}~\bibnamefont {Ma}}, \bibinfo {author} {\bibfnamefont {A.}~\bibnamefont {Arora}}, \bibinfo {author} {\bibfnamefont {G.}~\bibnamefont {Vignale}},\ and\ \bibinfo {author} {\bibfnamefont {J.~C.}\ \bibnamefont {Song}},\ }\bibfield  {title} {\bibinfo {title} {Anomalous skew-scattering nonlinear {Hall} effect and chiral photocurrents in {PT}-symmetric antiferromagnets},\ }\href@noop {} {\bibfield  {journal} {\bibinfo  {journal} {Phys. Rev. Lett.}\ }\textbf {\bibinfo {volume} {131}},\ \bibinfo {pages} {076601} (\bibinfo {year} {2023})}\BibitemShut {NoStop}%
\bibitem [{\citenamefont {Bouhon}\ \emph {et~al.}(2023)\citenamefont {Bouhon}, \citenamefont {Timmel},\ and\ \citenamefont {Slager}}]{bouhon2023quantum}%
  \BibitemOpen
  \bibfield  {author} {\bibinfo {author} {\bibfnamefont {A.}~\bibnamefont {Bouhon}}, \bibinfo {author} {\bibfnamefont {A.}~\bibnamefont {Timmel}},\ and\ \bibinfo {author} {\bibfnamefont {R.-J.}\ \bibnamefont {Slager}},\ }\bibfield  {title} {\bibinfo {title} {Quantum geometry beyond projective single bands},\ }\href@noop {} {\bibfield  {journal} {\bibinfo  {journal} {arXiv preprint arXiv:2303.02180}\ } (\bibinfo {year} {2023})}\BibitemShut {NoStop}%
\bibitem [{\citenamefont {Jankowski}\ and\ \citenamefont {Slager}(2024)}]{jankowski2024quantized}%
  \BibitemOpen
  \bibfield  {author} {\bibinfo {author} {\bibfnamefont {W.~J.}\ \bibnamefont {Jankowski}}\ and\ \bibinfo {author} {\bibfnamefont {R.-J.}\ \bibnamefont {Slager}},\ }\bibfield  {title} {\bibinfo {title} {Quantized integrated shift effect in multigap topological phases},\ }\href@noop {} {\bibfield  {journal} {\bibinfo  {journal} {Phys. Rev. Lett.}\ }\textbf {\bibinfo {volume} {133}},\ \bibinfo {pages} {186601} (\bibinfo {year} {2024})}\BibitemShut {NoStop}%
\bibitem [{\citenamefont {Jankowski}\ \emph {et~al.}(2025)\citenamefont {Jankowski}, \citenamefont {Morris}, \citenamefont {Bouhon}, \citenamefont {{\"U}nal},\ and\ \citenamefont {Slager}}]{jankowski2025optical}%
  \BibitemOpen
  \bibfield  {author} {\bibinfo {author} {\bibfnamefont {W.~J.}\ \bibnamefont {Jankowski}}, \bibinfo {author} {\bibfnamefont {A.~S.}\ \bibnamefont {Morris}}, \bibinfo {author} {\bibfnamefont {A.}~\bibnamefont {Bouhon}}, \bibinfo {author} {\bibfnamefont {F.~N.}\ \bibnamefont {{\"U}nal}},\ and\ \bibinfo {author} {\bibfnamefont {R.-J.}\ \bibnamefont {Slager}},\ }\bibfield  {title} {\bibinfo {title} {Optical manifestations and bounds of topological {Euler} class},\ }\href@noop {} {\bibfield  {journal} {\bibinfo  {journal} {Phys. Rev. B}\ }\textbf {\bibinfo {volume} {111}},\ \bibinfo {pages} {L081103} (\bibinfo {year} {2025})}\BibitemShut {NoStop}%
\bibitem [{\citenamefont {Avdoshkin}\ \emph {et~al.}(2024)\citenamefont {Avdoshkin}, \citenamefont {Mitscherling},\ and\ \citenamefont {Moore}}]{avdoshkin2024multi}%
  \BibitemOpen
  \bibfield  {author} {\bibinfo {author} {\bibfnamefont {A.}~\bibnamefont {Avdoshkin}}, \bibinfo {author} {\bibfnamefont {J.}~\bibnamefont {Mitscherling}},\ and\ \bibinfo {author} {\bibfnamefont {J.~E.}\ \bibnamefont {Moore}},\ }\bibfield  {title} {\bibinfo {title} {The multi-state geometry of shift current and polarization},\ }\href@noop {} {\bibfield  {journal} {\bibinfo  {journal} {arXiv preprint arXiv:2409.16358}\ } (\bibinfo {year} {2024})}\BibitemShut {NoStop}%
\bibitem [{\citenamefont {Sala}\ \emph {et~al.}(2024)\citenamefont {Sala}, \citenamefont {Mercaldo}, \citenamefont {Domi}, \citenamefont {Gariglio}, \citenamefont {Cuoco}, \citenamefont {Ortix},\ and\ \citenamefont {Caviglia}}]{sala2024quantum}%
  \BibitemOpen
  \bibfield  {author} {\bibinfo {author} {\bibfnamefont {G.}~\bibnamefont {Sala}}, \bibinfo {author} {\bibfnamefont {M.~T.}\ \bibnamefont {Mercaldo}}, \bibinfo {author} {\bibfnamefont {K.}~\bibnamefont {Domi}}, \bibinfo {author} {\bibfnamefont {S.}~\bibnamefont {Gariglio}}, \bibinfo {author} {\bibfnamefont {M.}~\bibnamefont {Cuoco}}, \bibinfo {author} {\bibfnamefont {C.}~\bibnamefont {Ortix}},\ and\ \bibinfo {author} {\bibfnamefont {A.~D.}\ \bibnamefont {Caviglia}},\ }\bibfield  {title} {\bibinfo {title} {The quantum metric of electrons with spin-momentum locking},\ }\href@noop {} {\bibfield  {journal} {\bibinfo  {journal} {arXiv preprint arXiv:2407.06659}\ } (\bibinfo {year} {2024})}\BibitemShut {NoStop}%
\bibitem [{\citenamefont {Gao}\ \emph {et~al.}(2015)\citenamefont {Gao}, \citenamefont {Yang},\ and\ \citenamefont {Niu}}]{gao2015geometrical}%
  \BibitemOpen
  \bibfield  {author} {\bibinfo {author} {\bibfnamefont {Y.}~\bibnamefont {Gao}}, \bibinfo {author} {\bibfnamefont {S.~A.}\ \bibnamefont {Yang}},\ and\ \bibinfo {author} {\bibfnamefont {Q.}~\bibnamefont {Niu}},\ }\bibfield  {title} {\bibinfo {title} {Geometrical effects in orbital magnetic susceptibility},\ }\href@noop {} {\bibfield  {journal} {\bibinfo  {journal} {Phys. Rev. B}\ }\textbf {\bibinfo {volume} {91}},\ \bibinfo {pages} {214405} (\bibinfo {year} {2015})}\BibitemShut {NoStop}%
\bibitem [{\citenamefont {Pi{\'e}chon}\ \emph {et~al.}(2016)\citenamefont {Pi{\'e}chon}, \citenamefont {Raoux}, \citenamefont {Fuchs},\ and\ \citenamefont {Montambaux}}]{piechon2016geometric}%
  \BibitemOpen
  \bibfield  {author} {\bibinfo {author} {\bibfnamefont {F.}~\bibnamefont {Pi{\'e}chon}}, \bibinfo {author} {\bibfnamefont {A.}~\bibnamefont {Raoux}}, \bibinfo {author} {\bibfnamefont {J.-N.}\ \bibnamefont {Fuchs}},\ and\ \bibinfo {author} {\bibfnamefont {G.}~\bibnamefont {Montambaux}},\ }\bibfield  {title} {\bibinfo {title} {Geometric orbital susceptibility: Quantum metric without {Berry} curvature},\ }\href@noop {} {\bibfield  {journal} {\bibinfo  {journal} {Phys. Rev. B}\ }\textbf {\bibinfo {volume} {94}},\ \bibinfo {pages} {134423} (\bibinfo {year} {2016})}\BibitemShut {NoStop}%
\bibitem [{\citenamefont {Julku}\ \emph {et~al.}(2016)\citenamefont {Julku}, \citenamefont {Peotta}, \citenamefont {Vanhala}, \citenamefont {Kim},\ and\ \citenamefont {T{\"o}rm{\"a}}}]{julku2016geometric}%
  \BibitemOpen
  \bibfield  {author} {\bibinfo {author} {\bibfnamefont {A.}~\bibnamefont {Julku}}, \bibinfo {author} {\bibfnamefont {S.}~\bibnamefont {Peotta}}, \bibinfo {author} {\bibfnamefont {T.~I.}\ \bibnamefont {Vanhala}}, \bibinfo {author} {\bibfnamefont {D.-H.}\ \bibnamefont {Kim}},\ and\ \bibinfo {author} {\bibfnamefont {P.}~\bibnamefont {T{\"o}rm{\"a}}},\ }\bibfield  {title} {\bibinfo {title} {Geometric origin of superfluidity in the lieb-lattice flat band},\ }\href@noop {} {\bibfield  {journal} {\bibinfo  {journal} {Phys. Rev. Lett.}\ }\textbf {\bibinfo {volume} {117}},\ \bibinfo {pages} {045303} (\bibinfo {year} {2016})}\BibitemShut {NoStop}%
\bibitem [{\citenamefont {Liang}\ \emph {et~al.}(2017)\citenamefont {Liang}, \citenamefont {Vanhala}, \citenamefont {Peotta}, \citenamefont {Siro}, \citenamefont {Harju},\ and\ \citenamefont {T{\"o}rm{\"a}}}]{liang2017band}%
  \BibitemOpen
  \bibfield  {author} {\bibinfo {author} {\bibfnamefont {L.}~\bibnamefont {Liang}}, \bibinfo {author} {\bibfnamefont {T.~I.}\ \bibnamefont {Vanhala}}, \bibinfo {author} {\bibfnamefont {S.}~\bibnamefont {Peotta}}, \bibinfo {author} {\bibfnamefont {T.}~\bibnamefont {Siro}}, \bibinfo {author} {\bibfnamefont {A.}~\bibnamefont {Harju}},\ and\ \bibinfo {author} {\bibfnamefont {P.}~\bibnamefont {T{\"o}rm{\"a}}},\ }\bibfield  {title} {\bibinfo {title} {Band geometry, {Berry} curvature, and superfluid weight},\ }\href@noop {} {\bibfield  {journal} {\bibinfo  {journal} {Phys. Rev. B}\ }\textbf {\bibinfo {volume} {95}},\ \bibinfo {pages} {024515} (\bibinfo {year} {2017})}\BibitemShut {NoStop}%
\bibitem [{\citenamefont {Hu}\ \emph {et~al.}(2019)\citenamefont {Hu}, \citenamefont {Hyart}, \citenamefont {Pikulin},\ and\ \citenamefont {Rossi}}]{hu2019geometric}%
  \BibitemOpen
  \bibfield  {author} {\bibinfo {author} {\bibfnamefont {X.}~\bibnamefont {Hu}}, \bibinfo {author} {\bibfnamefont {T.}~\bibnamefont {Hyart}}, \bibinfo {author} {\bibfnamefont {D.~I.}\ \bibnamefont {Pikulin}},\ and\ \bibinfo {author} {\bibfnamefont {E.}~\bibnamefont {Rossi}},\ }\bibfield  {title} {\bibinfo {title} {Geometric and conventional contribution to the superfluid weight in twisted bilayer graphene},\ }\href@noop {} {\bibfield  {journal} {\bibinfo  {journal} {Phys. Rev. Lett.}\ }\textbf {\bibinfo {volume} {123}},\ \bibinfo {pages} {237002} (\bibinfo {year} {2019})}\BibitemShut {NoStop}%
\bibitem [{\citenamefont {Xie}\ \emph {et~al.}(2020)\citenamefont {Xie}, \citenamefont {Song}, \citenamefont {Lian},\ and\ \citenamefont {Bernevig}}]{xie2020topology}%
  \BibitemOpen
  \bibfield  {author} {\bibinfo {author} {\bibfnamefont {F.}~\bibnamefont {Xie}}, \bibinfo {author} {\bibfnamefont {Z.}~\bibnamefont {Song}}, \bibinfo {author} {\bibfnamefont {B.}~\bibnamefont {Lian}},\ and\ \bibinfo {author} {\bibfnamefont {B.~A.}\ \bibnamefont {Bernevig}},\ }\bibfield  {title} {\bibinfo {title} {Topology-bounded superfluid weight in twisted bilayer graphene},\ }\href@noop {} {\bibfield  {journal} {\bibinfo  {journal} {Phys. Rev. Lett.}\ }\textbf {\bibinfo {volume} {124}},\ \bibinfo {pages} {167002} (\bibinfo {year} {2020})}\BibitemShut {NoStop}%
\bibitem [{\citenamefont {Julku}\ \emph {et~al.}(2020)\citenamefont {Julku}, \citenamefont {Peltonen}, \citenamefont {Liang}, \citenamefont {Heikkil{\"a}},\ and\ \citenamefont {T{\"o}rm{\"a}}}]{julku2020superfluid}%
  \BibitemOpen
  \bibfield  {author} {\bibinfo {author} {\bibfnamefont {A.}~\bibnamefont {Julku}}, \bibinfo {author} {\bibfnamefont {T.~J.}\ \bibnamefont {Peltonen}}, \bibinfo {author} {\bibfnamefont {L.}~\bibnamefont {Liang}}, \bibinfo {author} {\bibfnamefont {T.~T.}\ \bibnamefont {Heikkil{\"a}}},\ and\ \bibinfo {author} {\bibfnamefont {P.}~\bibnamefont {T{\"o}rm{\"a}}},\ }\bibfield  {title} {\bibinfo {title} {Superfluid weight and {Berezinskii-Kosterlitz-Thouless} transition temperature of twisted bilayer graphene},\ }\href@noop {} {\bibfield  {journal} {\bibinfo  {journal} {Phys. Rev. B}\ }\textbf {\bibinfo {volume} {101}},\ \bibinfo {pages} {060505} (\bibinfo {year} {2020})}\BibitemShut {NoStop}%
\bibitem [{\citenamefont {Rossi}(2021)}]{rossi2021quantum}%
  \BibitemOpen
  \bibfield  {author} {\bibinfo {author} {\bibfnamefont {E.}~\bibnamefont {Rossi}},\ }\bibfield  {title} {\bibinfo {title} {Quantum metric and correlated states in two-dimensional systems},\ }\href@noop {} {\bibfield  {journal} {\bibinfo  {journal} {Current Opinion in Solid State and Materials Science}\ }\textbf {\bibinfo {volume} {25}},\ \bibinfo {pages} {100952} (\bibinfo {year} {2021})}\BibitemShut {NoStop}%
\bibitem [{\citenamefont {T{\"o}rm{\"a}}\ \emph {et~al.}(2022)\citenamefont {T{\"o}rm{\"a}}, \citenamefont {Peotta},\ and\ \citenamefont {Bernevig}}]{torma2022superconductivity}%
  \BibitemOpen
  \bibfield  {author} {\bibinfo {author} {\bibfnamefont {P.}~\bibnamefont {T{\"o}rm{\"a}}}, \bibinfo {author} {\bibfnamefont {S.}~\bibnamefont {Peotta}},\ and\ \bibinfo {author} {\bibfnamefont {B.~A.}\ \bibnamefont {Bernevig}},\ }\bibfield  {title} {\bibinfo {title} {Superconductivity, superfluidity and quantum geometry in twisted multilayer systems},\ }\href@noop {} {\bibfield  {journal} {\bibinfo  {journal} {Nature Reviews Physics}\ }\textbf {\bibinfo {volume} {4}},\ \bibinfo {pages} {528} (\bibinfo {year} {2022})}\BibitemShut {NoStop}%
\bibitem [{\citenamefont {Tian}\ \emph {et~al.}(2023)\citenamefont {Tian}, \citenamefont {Gao}, \citenamefont {Zhang}, \citenamefont {Che}, \citenamefont {Xu}, \citenamefont {Cheung}, \citenamefont {Watanabe}, \citenamefont {Taniguchi}, \citenamefont {Randeria}, \citenamefont {Zhang} \emph {et~al.}}]{tian2023evidence}%
  \BibitemOpen
  \bibfield  {author} {\bibinfo {author} {\bibfnamefont {H.}~\bibnamefont {Tian}}, \bibinfo {author} {\bibfnamefont {X.}~\bibnamefont {Gao}}, \bibinfo {author} {\bibfnamefont {Y.}~\bibnamefont {Zhang}}, \bibinfo {author} {\bibfnamefont {S.}~\bibnamefont {Che}}, \bibinfo {author} {\bibfnamefont {T.}~\bibnamefont {Xu}}, \bibinfo {author} {\bibfnamefont {P.}~\bibnamefont {Cheung}}, \bibinfo {author} {\bibfnamefont {K.}~\bibnamefont {Watanabe}}, \bibinfo {author} {\bibfnamefont {T.}~\bibnamefont {Taniguchi}}, \bibinfo {author} {\bibfnamefont {M.}~\bibnamefont {Randeria}}, \bibinfo {author} {\bibfnamefont {F.}~\bibnamefont {Zhang}}, \emph {et~al.},\ }\bibfield  {title} {\bibinfo {title} {Evidence for {Dirac} flat band superconductivity enabled by quantum geometry},\ }\href@noop {} {\bibfield  {journal} {\bibinfo  {journal} {Nature}\ }\textbf {\bibinfo {volume} {614}},\ \bibinfo {pages} {440} (\bibinfo {year} {2023})}\BibitemShut {NoStop}%
\bibitem [{\citenamefont {Wang}\ \emph {et~al.}(2021)\citenamefont {Wang}, \citenamefont {Cano}, \citenamefont {Millis}, \citenamefont {Liu},\ and\ \citenamefont {Yang}}]{wang2021exact}%
  \BibitemOpen
  \bibfield  {author} {\bibinfo {author} {\bibfnamefont {J.}~\bibnamefont {Wang}}, \bibinfo {author} {\bibfnamefont {J.}~\bibnamefont {Cano}}, \bibinfo {author} {\bibfnamefont {A.~J.}\ \bibnamefont {Millis}}, \bibinfo {author} {\bibfnamefont {Z.}~\bibnamefont {Liu}},\ and\ \bibinfo {author} {\bibfnamefont {B.}~\bibnamefont {Yang}},\ }\bibfield  {title} {\bibinfo {title} {Exact landau level description of geometry and interaction in a flatband},\ }\href@noop {} {\bibfield  {journal} {\bibinfo  {journal} {Phys. Rev. Lett.}\ }\textbf {\bibinfo {volume} {127}},\ \bibinfo {pages} {246403} (\bibinfo {year} {2021})}\BibitemShut {NoStop}%
\bibitem [{\citenamefont {Ledwith}\ \emph {et~al.}(2023)\citenamefont {Ledwith}, \citenamefont {Vishwanath},\ and\ \citenamefont {Parker}}]{ledwith2023vortexability}%
  \BibitemOpen
  \bibfield  {author} {\bibinfo {author} {\bibfnamefont {P.~J.}\ \bibnamefont {Ledwith}}, \bibinfo {author} {\bibfnamefont {A.}~\bibnamefont {Vishwanath}},\ and\ \bibinfo {author} {\bibfnamefont {D.~E.}\ \bibnamefont {Parker}},\ }\bibfield  {title} {\bibinfo {title} {Vortexability: A unifying criterion for ideal fractional {Chern} insulators},\ }\href@noop {} {\bibfield  {journal} {\bibinfo  {journal} {Phys. Rev. B}\ }\textbf {\bibinfo {volume} {108}},\ \bibinfo {pages} {205144} (\bibinfo {year} {2023})}\BibitemShut {NoStop}%
\bibitem [{\citenamefont {Estienne}\ \emph {et~al.}(2023)\citenamefont {Estienne}, \citenamefont {Regnault},\ and\ \citenamefont {Cr{\'e}pel}}]{estienne2023ideal}%
  \BibitemOpen
  \bibfield  {author} {\bibinfo {author} {\bibfnamefont {B.}~\bibnamefont {Estienne}}, \bibinfo {author} {\bibfnamefont {N.}~\bibnamefont {Regnault}},\ and\ \bibinfo {author} {\bibfnamefont {V.}~\bibnamefont {Cr{\'e}pel}},\ }\bibfield  {title} {\bibinfo {title} {Ideal {Chern} bands as {Landau} levels in curved space},\ }\href@noop {} {\bibfield  {journal} {\bibinfo  {journal} {Phys. Rev. Res.}\ }\textbf {\bibinfo {volume} {5}},\ \bibinfo {pages} {L032048} (\bibinfo {year} {2023})}\BibitemShut {NoStop}%
\bibitem [{\citenamefont {Liu}\ \emph {et~al.}(2024)\citenamefont {Liu}, \citenamefont {Mera}, \citenamefont {Fujimoto}, \citenamefont {Ozawa},\ and\ \citenamefont {Wang}}]{liu2024theory}%
  \BibitemOpen
  \bibfield  {author} {\bibinfo {author} {\bibfnamefont {Z.}~\bibnamefont {Liu}}, \bibinfo {author} {\bibfnamefont {B.}~\bibnamefont {Mera}}, \bibinfo {author} {\bibfnamefont {M.}~\bibnamefont {Fujimoto}}, \bibinfo {author} {\bibfnamefont {T.}~\bibnamefont {Ozawa}},\ and\ \bibinfo {author} {\bibfnamefont {J.}~\bibnamefont {Wang}},\ }\bibfield  {title} {\bibinfo {title} {Theory of generalized {Landau} levels and implication for non-{Abelian} states},\ }\href@noop {} {\bibfield  {journal} {\bibinfo  {journal} {arXiv preprint arXiv:2405.14479}\ } (\bibinfo {year} {2024})}\BibitemShut {NoStop}%
\bibitem [{\citenamefont {Smith}\ \emph {et~al.}(2022)\citenamefont {Smith}, \citenamefont {Pullasseri},\ and\ \citenamefont {Srivastava}}]{smith2022momentum}%
  \BibitemOpen
  \bibfield  {author} {\bibinfo {author} {\bibfnamefont {T.~B.}\ \bibnamefont {Smith}}, \bibinfo {author} {\bibfnamefont {L.}~\bibnamefont {Pullasseri}},\ and\ \bibinfo {author} {\bibfnamefont {A.}~\bibnamefont {Srivastava}},\ }\bibfield  {title} {\bibinfo {title} {Momentum-space gravity from the quantum geometry and entropy of bloch electrons},\ }\href@noop {} {\bibfield  {journal} {\bibinfo  {journal} {Physical Review Research}\ }\textbf {\bibinfo {volume} {4}},\ \bibinfo {pages} {013217} (\bibinfo {year} {2022})}\BibitemShut {NoStop}%
\bibitem [{\citenamefont {Jain}\ \emph {et~al.}(2024)\citenamefont {Jain}, \citenamefont {Jankowski},\ and\ \citenamefont {Slager}}]{jain2024anomalous}%
  \BibitemOpen
  \bibfield  {author} {\bibinfo {author} {\bibfnamefont {A.}~\bibnamefont {Jain}}, \bibinfo {author} {\bibfnamefont {W.~J.}\ \bibnamefont {Jankowski}},\ and\ \bibinfo {author} {\bibfnamefont {R.-J.}\ \bibnamefont {Slager}},\ }\bibfield  {title} {\bibinfo {title} {Anomalous geometric transport signatures of topological euler class},\ }\href@noop {} {\bibfield  {journal} {\bibinfo  {journal} {arXiv preprint arXiv:2412.01810}\ } (\bibinfo {year} {2024})}\BibitemShut {NoStop}%
\bibitem [{\citenamefont {Mehraeen}(2025)}]{mehraeen2025quantum}%
  \BibitemOpen
  \bibfield  {author} {\bibinfo {author} {\bibfnamefont {M.}~\bibnamefont {Mehraeen}},\ }\bibfield  {title} {\bibinfo {title} {Quantum response theory and momentum-space gravity},\ }\href@noop {} {\bibfield  {journal} {\bibinfo  {journal} {arXiv preprint arXiv:2503.06160}\ } (\bibinfo {year} {2025})}\BibitemShut {NoStop}%
\bibitem [{\citenamefont {Jackiw}(1988)}]{jackiw1988three}%
  \BibitemOpen
  \bibfield  {author} {\bibinfo {author} {\bibfnamefont {R.}~\bibnamefont {Jackiw}},\ }\bibfield  {title} {\bibinfo {title} {Three elaborations on {Berry}’s connection, curvature and phase},\ }\href@noop {} {\bibfield  {journal} {\bibinfo  {journal} {International Journal of Modern Physics A}\ }\textbf {\bibinfo {volume} {3}},\ \bibinfo {pages} {285} (\bibinfo {year} {1988})}\BibitemShut {NoStop}%
\bibitem [{\citenamefont {Kolodrubetz}\ \emph {et~al.}(2017)\citenamefont {Kolodrubetz}, \citenamefont {Sels}, \citenamefont {Mehta},\ and\ \citenamefont {Polkovnikov}}]{kolodrubetz2017geometry}%
  \BibitemOpen
  \bibfield  {author} {\bibinfo {author} {\bibfnamefont {M.}~\bibnamefont {Kolodrubetz}}, \bibinfo {author} {\bibfnamefont {D.}~\bibnamefont {Sels}}, \bibinfo {author} {\bibfnamefont {P.}~\bibnamefont {Mehta}},\ and\ \bibinfo {author} {\bibfnamefont {A.}~\bibnamefont {Polkovnikov}},\ }\bibfield  {title} {\bibinfo {title} {Geometry and non-adiabatic response in quantum and classical systems},\ }\href@noop {} {\bibfield  {journal} {\bibinfo  {journal} {Physics Reports}\ }\textbf {\bibinfo {volume} {697}},\ \bibinfo {pages} {1} (\bibinfo {year} {2017})}\BibitemShut {NoStop}%
\bibitem [{\citenamefont {Goldhaber}(2005)}]{goldhaber2005newtonian}%
  \BibitemOpen
  \bibfield  {author} {\bibinfo {author} {\bibfnamefont {A.~S.}\ \bibnamefont {Goldhaber}},\ }\bibfield  {title} {\bibinfo {title} {Newtonian adiabatics unified},\ }\href@noop {} {\bibfield  {journal} {\bibinfo  {journal} {Phys. Rev. A}\ }\textbf {\bibinfo {volume} {71}},\ \bibinfo {pages} {062102} (\bibinfo {year} {2005})}\BibitemShut {NoStop}%
\bibitem [{\citenamefont {Requist}\ \emph {et~al.}(2016)\citenamefont {Requist}, \citenamefont {Tandetzky},\ and\ \citenamefont {Gross}}]{requist2016molecular}%
  \BibitemOpen
  \bibfield  {author} {\bibinfo {author} {\bibfnamefont {R.}~\bibnamefont {Requist}}, \bibinfo {author} {\bibfnamefont {F.}~\bibnamefont {Tandetzky}},\ and\ \bibinfo {author} {\bibfnamefont {E.}~\bibnamefont {Gross}},\ }\bibfield  {title} {\bibinfo {title} {Molecular geometric phase from the exact electron-nuclear factorization},\ }\href@noop {} {\bibfield  {journal} {\bibinfo  {journal} {Phys. Rev. A}\ }\textbf {\bibinfo {volume} {93}},\ \bibinfo {pages} {042108} (\bibinfo {year} {2016})}\BibitemShut {NoStop}%
\bibitem [{\citenamefont {Scherrer}\ \emph {et~al.}(2017)\citenamefont {Scherrer}, \citenamefont {Agostini}, \citenamefont {Sebastiani}, \citenamefont {Gross},\ and\ \citenamefont {Vuilleumier}}]{scherrer2017mass}%
  \BibitemOpen
  \bibfield  {author} {\bibinfo {author} {\bibfnamefont {A.}~\bibnamefont {Scherrer}}, \bibinfo {author} {\bibfnamefont {F.}~\bibnamefont {Agostini}}, \bibinfo {author} {\bibfnamefont {D.}~\bibnamefont {Sebastiani}}, \bibinfo {author} {\bibfnamefont {E.}~\bibnamefont {Gross}},\ and\ \bibinfo {author} {\bibfnamefont {R.}~\bibnamefont {Vuilleumier}},\ }\bibfield  {title} {\bibinfo {title} {On the mass of atoms in molecules: {Beyond the Born-Oppenheimer} approximation},\ }\href@noop {} {\bibfield  {journal} {\bibinfo  {journal} {Phys. Rev. X}\ }\textbf {\bibinfo {volume} {7}},\ \bibinfo {pages} {031035} (\bibinfo {year} {2017})}\BibitemShut {NoStop}%
\bibitem [{\citenamefont {Littlejohn}\ \emph {et~al.}(2023)\citenamefont {Littlejohn}, \citenamefont {Rawlinson},\ and\ \citenamefont {Subotnik}}]{littlejohn2023representation}%
  \BibitemOpen
  \bibfield  {author} {\bibinfo {author} {\bibfnamefont {R.}~\bibnamefont {Littlejohn}}, \bibinfo {author} {\bibfnamefont {J.}~\bibnamefont {Rawlinson}},\ and\ \bibinfo {author} {\bibfnamefont {J.}~\bibnamefont {Subotnik}},\ }\bibfield  {title} {\bibinfo {title} {Representation and conservation of angular momentum in the {Born--Oppenheimer} theory of polyatomic molecules},\ }\href@noop {} {\bibfield  {journal} {\bibinfo  {journal} {J. Chem. Phys.}\ }\textbf {\bibinfo {volume} {158}} (\bibinfo {year} {2023})}\BibitemShut {NoStop}%
\bibitem [{\citenamefont {Nelson}\ \emph {et~al.}(2020)\citenamefont {Nelson}, \citenamefont {White}, \citenamefont {Bjorgaard}, \citenamefont {Sifain}, \citenamefont {Zhang}, \citenamefont {Nebgen}, \citenamefont {Fernandez-Alberti}, \citenamefont {Mozyrsky}, \citenamefont {Roitberg},\ and\ \citenamefont {Tretiak}}]{nelson2020non}%
  \BibitemOpen
  \bibfield  {author} {\bibinfo {author} {\bibfnamefont {T.~R.}\ \bibnamefont {Nelson}}, \bibinfo {author} {\bibfnamefont {A.~J.}\ \bibnamefont {White}}, \bibinfo {author} {\bibfnamefont {J.~A.}\ \bibnamefont {Bjorgaard}}, \bibinfo {author} {\bibfnamefont {A.~E.}\ \bibnamefont {Sifain}}, \bibinfo {author} {\bibfnamefont {Y.}~\bibnamefont {Zhang}}, \bibinfo {author} {\bibfnamefont {B.}~\bibnamefont {Nebgen}}, \bibinfo {author} {\bibfnamefont {S.}~\bibnamefont {Fernandez-Alberti}}, \bibinfo {author} {\bibfnamefont {D.}~\bibnamefont {Mozyrsky}}, \bibinfo {author} {\bibfnamefont {A.~E.}\ \bibnamefont {Roitberg}},\ and\ \bibinfo {author} {\bibfnamefont {S.}~\bibnamefont {Tretiak}},\ }\bibfield  {title} {\bibinfo {title} {Non-adiabatic excited-state molecular dynamics: Theory and applications for modeling photophysics in extended molecular materials},\ }\href@noop {} {\bibfield  {journal} {\bibinfo  {journal} {Chem. Rev.}\ }\textbf {\bibinfo {volume} {120}},\ \bibinfo {pages} {2215} (\bibinfo {year}
  {2020})}\BibitemShut {NoStop}%
\bibitem [{\citenamefont {Allahverdyan}\ and\ \citenamefont {Mehmani}(2009)}]{allahverdyan2009post}%
  \BibitemOpen
  \bibfield  {author} {\bibinfo {author} {\bibfnamefont {A.}~\bibnamefont {Allahverdyan}}\ and\ \bibinfo {author} {\bibfnamefont {B.}~\bibnamefont {Mehmani}},\ }\bibfield  {title} {\bibinfo {title} {Post-adiabatic forces and lagrangians with higher-order derivatives},\ }\href@noop {} {\bibfield  {journal} {\bibinfo  {journal} {arXiv preprint arXiv:0905.1596}\ } (\bibinfo {year} {2009})}\BibitemShut {NoStop}%
\bibitem [{\citenamefont {Het{\'e}nyi}\ and\ \citenamefont {L{\'e}vay}(2023)}]{hetenyi2023fluctuations}%
  \BibitemOpen
  \bibfield  {author} {\bibinfo {author} {\bibfnamefont {B.}~\bibnamefont {Het{\'e}nyi}}\ and\ \bibinfo {author} {\bibfnamefont {P.}~\bibnamefont {L{\'e}vay}},\ }\bibfield  {title} {\bibinfo {title} {Fluctuations, uncertainty relations, and the geometry of quantum state manifolds},\ }\href@noop {} {\bibfield  {journal} {\bibinfo  {journal} {Phys. Rev. A}\ }\textbf {\bibinfo {volume} {108}},\ \bibinfo {pages} {032218} (\bibinfo {year} {2023})}\BibitemShut {NoStop}%
\bibitem [{\citenamefont {Sundaram}\ and\ \citenamefont {Niu}(1999)}]{sundaram1999wave}%
  \BibitemOpen
  \bibfield  {author} {\bibinfo {author} {\bibfnamefont {G.}~\bibnamefont {Sundaram}}\ and\ \bibinfo {author} {\bibfnamefont {Q.}~\bibnamefont {Niu}},\ }\bibfield  {title} {\bibinfo {title} {Wave-packet dynamics in slowly perturbed crystals: Gradient corrections and {Berry}-phase effects},\ }\href@noop {} {\bibfield  {journal} {\bibinfo  {journal} {Phys. Rev. B}\ }\textbf {\bibinfo {volume} {59}},\ \bibinfo {pages} {14915} (\bibinfo {year} {1999})}\BibitemShut {NoStop}%
\bibitem [{SM()}]{SM}%
  \BibitemOpen
  \bibinfo {note} {{See Supplemental Material [url] for the detailed derivations of the effective Lagrangian, equations of motion, effects of electric field, canonical quantization, flat-band nonadiabatic metric, and equations of motion of two-particle dynamics. Reference~[\onlinecite{ozawa2021relations}] is cited.}}\BibitemShut {Stop}%
\bibitem [{\citenamefont {{Dirac}}(1930)}]{dirac1930note}%
  \BibitemOpen
  \bibfield  {author} {\bibinfo {author} {\bibfnamefont {P.~A.}\ \bibnamefont {{Dirac}}},\ }\bibfield  {title} {\bibinfo {title} {Note on exchange phenomena in the thomas atom},\ }in\ \href@noop {} {\emph {\bibinfo {booktitle} {Mathematical proceedings of the Cambridge philosophical society}}},\ Vol.~\bibinfo {volume} {26}\ (\bibinfo {organization} {Cambridge University Press},\ \bibinfo {year} {1930})\ pp.\ \bibinfo {pages} {376--385}\BibitemShut {NoStop}%
\bibitem [{\citenamefont {Raab}(2000)}]{raab2000dirac}%
  \BibitemOpen
  \bibfield  {author} {\bibinfo {author} {\bibfnamefont {A.}~\bibnamefont {Raab}},\ }\bibfield  {title} {\bibinfo {title} {On the {Dirac}--{Frenkel/McLachlan} variational principle},\ }\href@noop {} {\bibfield  {journal} {\bibinfo  {journal} {Chem. Phys. Lett.}\ }\textbf {\bibinfo {volume} {319}},\ \bibinfo {pages} {674} (\bibinfo {year} {2000})}\BibitemShut {NoStop}%
\bibitem [{\citenamefont {Ren}\ and\ \citenamefont {Barrero}(2025)}]{ren2025nonadiabatic}%
  \BibitemOpen
  \bibfield  {author} {\bibinfo {author} {\bibfnamefont {Y.}~\bibnamefont {Ren}}\ and\ \bibinfo {author} {\bibfnamefont {M.}~\bibnamefont {Barrero}},\ }\bibfield  {title} {\bibinfo {title} {Nonadiabatic wave-packet dynamics: Nonadiabatic metric, quantum geometry, and analogue gravity},\ }\href@noop {} {\bibfield  {journal} {\bibinfo  {journal} {arXiv preprint arXiv:2509.00166}\ } (\bibinfo {year} {2025})}\BibitemShut {NoStop}%
\bibitem [{\citenamefont {Nakahara}(2018)}]{nakahara2018geometry}%
  \BibitemOpen
  \bibfield  {author} {\bibinfo {author} {\bibfnamefont {M.}~\bibnamefont {Nakahara}},\ }\href@noop {} {\emph {\bibinfo {title} {Geometry, topology and physics}}}\ (\bibinfo  {publisher} {CRC press},\ \bibinfo {year} {2018})\BibitemShut {NoStop}%
\bibitem [{\citenamefont {Komissarov}\ \emph {et~al.}(2024)\citenamefont {Komissarov}, \citenamefont {Holder},\ and\ \citenamefont {Queiroz}}]{komissarov2024quantum}%
  \BibitemOpen
  \bibfield  {author} {\bibinfo {author} {\bibfnamefont {I.}~\bibnamefont {Komissarov}}, \bibinfo {author} {\bibfnamefont {T.}~\bibnamefont {Holder}},\ and\ \bibinfo {author} {\bibfnamefont {R.}~\bibnamefont {Queiroz}},\ }\bibfield  {title} {\bibinfo {title} {The quantum geometric origin of capacitance in insulators},\ }\href@noop {} {\bibfield  {journal} {\bibinfo  {journal} {Nat. Commun.}\ }\textbf {\bibinfo {volume} {15}},\ \bibinfo {pages} {4621} (\bibinfo {year} {2024})}\BibitemShut {NoStop}%
\bibitem [{\citenamefont {Jones}\ and\ \citenamefont {Zener}(1934)}]{jones1934general}%
  \BibitemOpen
  \bibfield  {author} {\bibinfo {author} {\bibfnamefont {H.}~\bibnamefont {Jones}}\ and\ \bibinfo {author} {\bibfnamefont {C.}~\bibnamefont {Zener}},\ }\bibfield  {title} {\bibinfo {title} {The general proof of certain fundamental equations in the theory of metallic conduction},\ }\href@noop {} {\bibfield  {journal} {\bibinfo  {journal} {Proceedings of the Royal Society of London. Series A, Containing Papers of a Mathematical and Physical Character}\ }\textbf {\bibinfo {volume} {144}},\ \bibinfo {pages} {101} (\bibinfo {year} {1934})}\BibitemShut {NoStop}%
\bibitem [{\citenamefont {Chang}\ and\ \citenamefont {Niu}(1995)}]{chang1995berry}%
  \BibitemOpen
  \bibfield  {author} {\bibinfo {author} {\bibfnamefont {M.-C.}\ \bibnamefont {Chang}}\ and\ \bibinfo {author} {\bibfnamefont {Q.}~\bibnamefont {Niu}},\ }\bibfield  {title} {\bibinfo {title} {{Berry} phase, hyperorbits, and the {Hofstadter} spectrum},\ }\href@noop {} {\bibfield  {journal} {\bibinfo  {journal} {Phys. Rev. Lett.}\ }\textbf {\bibinfo {volume} {75}},\ \bibinfo {pages} {1348} (\bibinfo {year} {1995})}\BibitemShut {NoStop}%
\bibitem [{\citenamefont {Chang}\ and\ \citenamefont {Niu}(1996)}]{chang1996berry}%
  \BibitemOpen
  \bibfield  {author} {\bibinfo {author} {\bibfnamefont {M.-C.}\ \bibnamefont {Chang}}\ and\ \bibinfo {author} {\bibfnamefont {Q.}~\bibnamefont {Niu}},\ }\bibfield  {title} {\bibinfo {title} {{Berry} phase, hyperorbits, and the {Hofstadter} spectrum: Semiclassical dynamics in magnetic {Bloch} bands},\ }\href@noop {} {\bibfield  {journal} {\bibinfo  {journal} {Phys. Rev. B}\ }\textbf {\bibinfo {volume} {53}},\ \bibinfo {pages} {7010} (\bibinfo {year} {1996})}\BibitemShut {NoStop}%
\bibitem [{\citenamefont {Liu}\ \emph {et~al.}(2022)\citenamefont {Liu}, \citenamefont {Zhao}, \citenamefont {Huang}, \citenamefont {Feng}, \citenamefont {Xiao}, \citenamefont {Wu}, \citenamefont {Lai}, \citenamefont {Gao},\ and\ \citenamefont {Yang}}]{liu2022berry}%
  \BibitemOpen
  \bibfield  {author} {\bibinfo {author} {\bibfnamefont {H.}~\bibnamefont {Liu}}, \bibinfo {author} {\bibfnamefont {J.}~\bibnamefont {Zhao}}, \bibinfo {author} {\bibfnamefont {Y.-X.}\ \bibnamefont {Huang}}, \bibinfo {author} {\bibfnamefont {X.}~\bibnamefont {Feng}}, \bibinfo {author} {\bibfnamefont {C.}~\bibnamefont {Xiao}}, \bibinfo {author} {\bibfnamefont {W.}~\bibnamefont {Wu}}, \bibinfo {author} {\bibfnamefont {S.}~\bibnamefont {Lai}}, \bibinfo {author} {\bibfnamefont {W.-b.}\ \bibnamefont {Gao}},\ and\ \bibinfo {author} {\bibfnamefont {S.~A.}\ \bibnamefont {Yang}},\ }\bibfield  {title} {\bibinfo {title} {Berry connection polarizability tensor and third-order hall effect},\ }\href@noop {} {\bibfield  {journal} {\bibinfo  {journal} {Physical Review B}\ }\textbf {\bibinfo {volume} {105}},\ \bibinfo {pages} {045118} (\bibinfo {year} {2022})}\BibitemShut {NoStop}%
\bibitem [{\citenamefont {Resta}(2025)}]{resta2025nonadiabatic}%
  \BibitemOpen
  \bibfield  {author} {\bibinfo {author} {\bibfnamefont {R.}~\bibnamefont {Resta}},\ }\bibfield  {title} {\bibinfo {title} {Nonadiabatic quantum geometry and optical conductivity},\ }\href@noop {} {\bibfield  {journal} {\bibinfo  {journal} {Physical Review B}\ }\textbf {\bibinfo {volume} {111}},\ \bibinfo {pages} {205107} (\bibinfo {year} {2025})}\BibitemShut {NoStop}%
\bibitem [{\citenamefont {Oka}\ and\ \citenamefont {Kitamura}(2019)}]{oka2019floquet}%
  \BibitemOpen
  \bibfield  {author} {\bibinfo {author} {\bibfnamefont {T.}~\bibnamefont {Oka}}\ and\ \bibinfo {author} {\bibfnamefont {S.}~\bibnamefont {Kitamura}},\ }\bibfield  {title} {\bibinfo {title} {Floquet engineering of quantum materials},\ }\href@noop {} {\bibfield  {journal} {\bibinfo  {journal} {Annu. Rev. Condens. Matter Phys.}\ }\textbf {\bibinfo {volume} {10}},\ \bibinfo {pages} {387} (\bibinfo {year} {2019})}\BibitemShut {NoStop}%
\bibitem [{\citenamefont {Zhou}\ \emph {et~al.}(2023)\citenamefont {Zhou}, \citenamefont {Bao}, \citenamefont {Fan}, \citenamefont {Wang}, \citenamefont {Zhong}, \citenamefont {Zhang}, \citenamefont {Tang}, \citenamefont {Duan},\ and\ \citenamefont {Zhou}}]{zhou2023floquet}%
  \BibitemOpen
  \bibfield  {author} {\bibinfo {author} {\bibfnamefont {S.}~\bibnamefont {Zhou}}, \bibinfo {author} {\bibfnamefont {C.}~\bibnamefont {Bao}}, \bibinfo {author} {\bibfnamefont {B.}~\bibnamefont {Fan}}, \bibinfo {author} {\bibfnamefont {F.}~\bibnamefont {Wang}}, \bibinfo {author} {\bibfnamefont {H.}~\bibnamefont {Zhong}}, \bibinfo {author} {\bibfnamefont {H.}~\bibnamefont {Zhang}}, \bibinfo {author} {\bibfnamefont {P.}~\bibnamefont {Tang}}, \bibinfo {author} {\bibfnamefont {W.}~\bibnamefont {Duan}},\ and\ \bibinfo {author} {\bibfnamefont {S.}~\bibnamefont {Zhou}},\ }\bibfield  {title} {\bibinfo {title} {Floquet engineering of black phosphorus upon below-gap pumping},\ }\href@noop {} {\bibfield  {journal} {\bibinfo  {journal} {Phys. Rev. Lett.}\ }\textbf {\bibinfo {volume} {131}},\ \bibinfo {pages} {116401} (\bibinfo {year} {2023})}\BibitemShut {NoStop}%
\bibitem [{\citenamefont {Zhan}\ \emph {et~al.}(2024)\citenamefont {Zhan}, \citenamefont {Chen}, \citenamefont {Ning}, \citenamefont {Ma}, \citenamefont {Wang}, \citenamefont {Xu},\ and\ \citenamefont {Wang}}]{zhan2024perspective}%
  \BibitemOpen
  \bibfield  {author} {\bibinfo {author} {\bibfnamefont {F.}~\bibnamefont {Zhan}}, \bibinfo {author} {\bibfnamefont {R.}~\bibnamefont {Chen}}, \bibinfo {author} {\bibfnamefont {Z.}~\bibnamefont {Ning}}, \bibinfo {author} {\bibfnamefont {D.-S.}\ \bibnamefont {Ma}}, \bibinfo {author} {\bibfnamefont {Z.}~\bibnamefont {Wang}}, \bibinfo {author} {\bibfnamefont {D.-H.}\ \bibnamefont {Xu}},\ and\ \bibinfo {author} {\bibfnamefont {R.}~\bibnamefont {Wang}},\ }\bibfield  {title} {\bibinfo {title} {Perspective: {Floquet} engineering topological states from effective models towards realistic materials},\ }\href@noop {} {\bibfield  {journal} {\bibinfo  {journal} {Quantum Frontiers}\ }\textbf {\bibinfo {volume} {3}},\ \bibinfo {pages} {21} (\bibinfo {year} {2024})}\BibitemShut {NoStop}%
\bibitem [{\citenamefont {Choi}\ \emph {et~al.}(2025)\citenamefont {Choi}, \citenamefont {Mogi}, \citenamefont {De~Giovannini}, \citenamefont {Azoury}, \citenamefont {Lv}, \citenamefont {Su}, \citenamefont {H{\"u}bener}, \citenamefont {Rubio},\ and\ \citenamefont {Gedik}}]{choi2025observation}%
  \BibitemOpen
  \bibfield  {author} {\bibinfo {author} {\bibfnamefont {D.}~\bibnamefont {Choi}}, \bibinfo {author} {\bibfnamefont {M.}~\bibnamefont {Mogi}}, \bibinfo {author} {\bibfnamefont {U.}~\bibnamefont {De~Giovannini}}, \bibinfo {author} {\bibfnamefont {D.}~\bibnamefont {Azoury}}, \bibinfo {author} {\bibfnamefont {B.}~\bibnamefont {Lv}}, \bibinfo {author} {\bibfnamefont {Y.}~\bibnamefont {Su}}, \bibinfo {author} {\bibfnamefont {H.}~\bibnamefont {H{\"u}bener}}, \bibinfo {author} {\bibfnamefont {A.}~\bibnamefont {Rubio}},\ and\ \bibinfo {author} {\bibfnamefont {N.}~\bibnamefont {Gedik}},\ }\bibfield  {title} {\bibinfo {title} {Observation of {Floquet}--{Bloch} states in monolayer graphene},\ }\href@noop {} {\bibfield  {journal} {\bibinfo  {journal} {Nat. Phys.}\ ,\ \bibinfo {pages} {1}} (\bibinfo {year} {2025})}\BibitemShut {NoStop}%
\bibitem [{\citenamefont {Nakata}\ \emph {et~al.}(2017)\citenamefont {Nakata}, \citenamefont {Kim}, \citenamefont {Klinovaja},\ and\ \citenamefont {Loss}}]{nakata2017magnonic}%
  \BibitemOpen
  \bibfield  {author} {\bibinfo {author} {\bibfnamefont {K.}~\bibnamefont {Nakata}}, \bibinfo {author} {\bibfnamefont {S.~K.}\ \bibnamefont {Kim}}, \bibinfo {author} {\bibfnamefont {J.}~\bibnamefont {Klinovaja}},\ and\ \bibinfo {author} {\bibfnamefont {D.}~\bibnamefont {Loss}},\ }\bibfield  {title} {\bibinfo {title} {Magnonic topological insulators in antiferromagnets},\ }\href@noop {} {\bibfield  {journal} {\bibinfo  {journal} {Phys. Rev. B}\ }\textbf {\bibinfo {volume} {96}},\ \bibinfo {pages} {224414} (\bibinfo {year} {2017})}\BibitemShut {NoStop}%
\bibitem [{\citenamefont {Li}\ \emph {et~al.}(2021)\citenamefont {Li}, \citenamefont {Cao},\ and\ \citenamefont {Yan}}]{li2021topological}%
  \BibitemOpen
  \bibfield  {author} {\bibinfo {author} {\bibfnamefont {Z.-X.}\ \bibnamefont {Li}}, \bibinfo {author} {\bibfnamefont {Y.}~\bibnamefont {Cao}},\ and\ \bibinfo {author} {\bibfnamefont {P.}~\bibnamefont {Yan}},\ }\bibfield  {title} {\bibinfo {title} {Topological insulators and semimetals in classical magnetic systems},\ }\href@noop {} {\bibfield  {journal} {\bibinfo  {journal} {Phys. Rep.}\ }\textbf {\bibinfo {volume} {915}},\ \bibinfo {pages} {1} (\bibinfo {year} {2021})}\BibitemShut {NoStop}%
\bibitem [{\citenamefont {Tang}\ and\ \citenamefont {Cheng}(2024)}]{tang2024lossless}%
  \BibitemOpen
  \bibfield  {author} {\bibinfo {author} {\bibfnamefont {J.}~\bibnamefont {Tang}}\ and\ \bibinfo {author} {\bibfnamefont {R.}~\bibnamefont {Cheng}},\ }\bibfield  {title} {\bibinfo {title} {Lossless spin-orbit torque in antiferromagnetic topological insulator mnbi 2 te 4},\ }\href@noop {} {\bibfield  {journal} {\bibinfo  {journal} {Phys. Rev. Lett.}\ }\textbf {\bibinfo {volume} {132}},\ \bibinfo {pages} {136701} (\bibinfo {year} {2024})}\BibitemShut {NoStop}%
\bibitem [{\citenamefont {Cheng}\ \emph {et~al.}(2020)\citenamefont {Cheng}, \citenamefont {Schumann}, \citenamefont {Wang}, \citenamefont {Zhang}, \citenamefont {Barbalas}, \citenamefont {Stemmer},\ and\ \citenamefont {Armitage}}]{cheng2020large}%
  \BibitemOpen
  \bibfield  {author} {\bibinfo {author} {\bibfnamefont {B.}~\bibnamefont {Cheng}}, \bibinfo {author} {\bibfnamefont {T.}~\bibnamefont {Schumann}}, \bibinfo {author} {\bibfnamefont {Y.}~\bibnamefont {Wang}}, \bibinfo {author} {\bibfnamefont {X.}~\bibnamefont {Zhang}}, \bibinfo {author} {\bibfnamefont {D.}~\bibnamefont {Barbalas}}, \bibinfo {author} {\bibfnamefont {S.}~\bibnamefont {Stemmer}},\ and\ \bibinfo {author} {\bibfnamefont {N.}~\bibnamefont {Armitage}},\ }\bibfield  {title} {\bibinfo {title} {A large effective phonon magnetic moment in a {Dirac} semimetal},\ }\href@noop {} {\bibfield  {journal} {\bibinfo  {journal} {Nano Lett.}\ }\textbf {\bibinfo {volume} {20}},\ \bibinfo {pages} {5991} (\bibinfo {year} {2020})}\BibitemShut {NoStop}%
\bibitem [{\citenamefont {Basini}\ \emph {et~al.}(2024)\citenamefont {Basini}, \citenamefont {Pancaldi}, \citenamefont {Wehinger}, \citenamefont {Udina}, \citenamefont {Unikandanunni}, \citenamefont {Tadano}, \citenamefont {Hoffmann}, \citenamefont {Balatsky},\ and\ \citenamefont {Bonetti}}]{basini2024terahertz}%
  \BibitemOpen
  \bibfield  {author} {\bibinfo {author} {\bibfnamefont {M.}~\bibnamefont {Basini}}, \bibinfo {author} {\bibfnamefont {M.}~\bibnamefont {Pancaldi}}, \bibinfo {author} {\bibfnamefont {B.}~\bibnamefont {Wehinger}}, \bibinfo {author} {\bibfnamefont {M.}~\bibnamefont {Udina}}, \bibinfo {author} {\bibfnamefont {V.}~\bibnamefont {Unikandanunni}}, \bibinfo {author} {\bibfnamefont {T.}~\bibnamefont {Tadano}}, \bibinfo {author} {\bibfnamefont {M.~C.}\ \bibnamefont {Hoffmann}}, \bibinfo {author} {\bibfnamefont {A.~V.}\ \bibnamefont {Balatsky}},\ and\ \bibinfo {author} {\bibfnamefont {S.}~\bibnamefont {Bonetti}},\ }\bibfield  {title} {\bibinfo {title} {Terahertz electric-field-driven dynamical multiferroicity in {SrTiO$_3$}},\ }\href@noop {} {\bibfield  {journal} {\bibinfo  {journal} {Nature}\ }\textbf {\bibinfo {volume} {628}},\ \bibinfo {pages} {534} (\bibinfo {year} {2024})}\BibitemShut {NoStop}%
\bibitem [{\citenamefont {Xu}\ and\ \citenamefont {Zong}(2025)}]{xu2025time}%
  \BibitemOpen
  \bibfield  {author} {\bibinfo {author} {\bibfnamefont {C.}~\bibnamefont {Xu}}\ and\ \bibinfo {author} {\bibfnamefont {A.}~\bibnamefont {Zong}},\ }\bibfield  {title} {\bibinfo {title} {Time-domain study of coupled collective excitations in quantum materials},\ }\href@noop {} {\bibfield  {journal} {\bibinfo  {journal} {npj Quantum Mater.}\ }\textbf {\bibinfo {volume} {10}},\ \bibinfo {pages} {21} (\bibinfo {year} {2025})}\BibitemShut {NoStop}%
\bibitem [{\citenamefont {Bhalla}\ \emph {et~al.}(2022)\citenamefont {Bhalla}, \citenamefont {Das}, \citenamefont {Culcer},\ and\ \citenamefont {Agarwal}}]{bhalla2022resonant}%
  \BibitemOpen
  \bibfield  {author} {\bibinfo {author} {\bibfnamefont {P.}~\bibnamefont {Bhalla}}, \bibinfo {author} {\bibfnamefont {K.}~\bibnamefont {Das}}, \bibinfo {author} {\bibfnamefont {D.}~\bibnamefont {Culcer}},\ and\ \bibinfo {author} {\bibfnamefont {A.}~\bibnamefont {Agarwal}},\ }\bibfield  {title} {\bibinfo {title} {Resonant second-harmonic generation as a probe of quantum geometry},\ }\href@noop {} {\bibfield  {journal} {\bibinfo  {journal} {Physical review letters}\ }\textbf {\bibinfo {volume} {129}},\ \bibinfo {pages} {227401} (\bibinfo {year} {2022})}\BibitemShut {NoStop}%
\bibitem [{\citenamefont {Volovik}(2003)}]{volovik2003universe}%
  \BibitemOpen
  \bibfield  {author} {\bibinfo {author} {\bibfnamefont {G.~E.}\ \bibnamefont {Volovik}},\ }\href@noop {} {\emph {\bibinfo {title} {The universe in a helium droplet}}},\ Vol.\ \bibinfo {volume} {117}\ (\bibinfo  {publisher} {OUP Oxford},\ \bibinfo {year} {2003})\BibitemShut {NoStop}%
\bibitem [{\citenamefont {Chang}\ and\ \citenamefont {Niu}(2008)}]{chang2008berry}%
  \BibitemOpen
  \bibfield  {author} {\bibinfo {author} {\bibfnamefont {M.-C.}\ \bibnamefont {Chang}}\ and\ \bibinfo {author} {\bibfnamefont {Q.}~\bibnamefont {Niu}},\ }\bibfield  {title} {\bibinfo {title} {{Berry} curvature, orbital moment, and effective quantum theory of electrons inelectromagnetic fields},\ }\href@noop {} {\bibfield  {journal} {\bibinfo  {journal} {Journal of Physics: Condensed Matter}\ }\textbf {\bibinfo {volume} {20}},\ \bibinfo {pages} {193202} (\bibinfo {year} {2008})}\BibitemShut {NoStop}%
\bibitem [{\citenamefont {Zhou}\ \emph {et~al.}(2015)\citenamefont {Zhou}, \citenamefont {Shan}, \citenamefont {Yao},\ and\ \citenamefont {Xiao}}]{zhou2015berry}%
  \BibitemOpen
  \bibfield  {author} {\bibinfo {author} {\bibfnamefont {J.}~\bibnamefont {Zhou}}, \bibinfo {author} {\bibfnamefont {W.-Y.}\ \bibnamefont {Shan}}, \bibinfo {author} {\bibfnamefont {W.}~\bibnamefont {Yao}},\ and\ \bibinfo {author} {\bibfnamefont {D.}~\bibnamefont {Xiao}},\ }\bibfield  {title} {\bibinfo {title} {{Berry} phase modification to the energy spectrum of excitons},\ }\href@noop {} {\bibfield  {journal} {\bibinfo  {journal} {Phys. Rev. Lett.}\ }\textbf {\bibinfo {volume} {115}},\ \bibinfo {pages} {166803} (\bibinfo {year} {2015})}\BibitemShut {NoStop}%
\bibitem [{\citenamefont {Shi}\ and\ \citenamefont {Ji}(2018)}]{shi2018dynamics}%
  \BibitemOpen
  \bibfield  {author} {\bibinfo {author} {\bibfnamefont {J.}~\bibnamefont {Shi}}\ and\ \bibinfo {author} {\bibfnamefont {W.}~\bibnamefont {Ji}},\ }\bibfield  {title} {\bibinfo {title} {Dynamics of the {Wigner} crystal of composite particles},\ }\href@noop {} {\bibfield  {journal} {\bibinfo  {journal} {Phys. Rev. B}\ }\textbf {\bibinfo {volume} {97}},\ \bibinfo {pages} {125133} (\bibinfo {year} {2018})}\BibitemShut {NoStop}%
\bibitem [{\citenamefont {Srivastava}\ and\ \citenamefont {Imamo{\u{g}}lu}(2015)}]{srivastava2015signatures}%
  \BibitemOpen
  \bibfield  {author} {\bibinfo {author} {\bibfnamefont {A.}~\bibnamefont {Srivastava}}\ and\ \bibinfo {author} {\bibfnamefont {A.}~\bibnamefont {Imamo{\u{g}}lu}},\ }\bibfield  {title} {\bibinfo {title} {{Signatures of Bloch-band geometry on excitons: nonhydrogenic spectra in transition-metal dichalcogenides}},\ }\href@noop {} {\bibfield  {journal} {\bibinfo  {journal} {Phys. Rev. Lett.}\ }\textbf {\bibinfo {volume} {115}},\ \bibinfo {pages} {166802} (\bibinfo {year} {2015})}\BibitemShut {NoStop}%
\bibitem [{\citenamefont {Skinner}(2016)}]{skinner2016interlayer}%
  \BibitemOpen
  \bibfield  {author} {\bibinfo {author} {\bibfnamefont {B.}~\bibnamefont {Skinner}},\ }\bibfield  {title} {\bibinfo {title} {Interlayer excitons with tunable dispersion relation},\ }\href@noop {} {\bibfield  {journal} {\bibinfo  {journal} {Phys. Rev. B}\ }\textbf {\bibinfo {volume} {93}},\ \bibinfo {pages} {235110} (\bibinfo {year} {2016})}\BibitemShut {NoStop}%
\bibitem [{\citenamefont {Wu}\ \emph {et~al.}(2017)\citenamefont {Wu}, \citenamefont {Lovorn},\ and\ \citenamefont {MacDonald}}]{wu2017topological}%
  \BibitemOpen
  \bibfield  {author} {\bibinfo {author} {\bibfnamefont {F.}~\bibnamefont {Wu}}, \bibinfo {author} {\bibfnamefont {T.}~\bibnamefont {Lovorn}},\ and\ \bibinfo {author} {\bibfnamefont {A.~H.}\ \bibnamefont {MacDonald}},\ }\bibfield  {title} {\bibinfo {title} {Topological exciton bands in moir{\'e} heterojunctions},\ }\href@noop {} {\bibfield  {journal} {\bibinfo  {journal} {Phys. Rev. Lett.}\ }\textbf {\bibinfo {volume} {118}},\ \bibinfo {pages} {147401} (\bibinfo {year} {2017})}\BibitemShut {NoStop}%
\bibitem [{\citenamefont {Joy}\ \emph {et~al.}(2025)\citenamefont {Joy}, \citenamefont {Levitov},\ and\ \citenamefont {Skinner}}]{joy2025chiral}%
  \BibitemOpen
  \bibfield  {author} {\bibinfo {author} {\bibfnamefont {S.}~\bibnamefont {Joy}}, \bibinfo {author} {\bibfnamefont {L.}~\bibnamefont {Levitov}},\ and\ \bibinfo {author} {\bibfnamefont {B.}~\bibnamefont {Skinner}},\ }\bibfield  {title} {\bibinfo {title} {{Chiral Wigner crystal phases induced by Berry curvature}},\ }\href@noop {} {\bibfield  {journal} {\bibinfo  {journal} {arXiv preprint arXiv:2507.22121}\ } (\bibinfo {year} {2025})}\BibitemShut {NoStop}%
\bibitem [{\citenamefont {Price}\ \emph {et~al.}(2014)\citenamefont {Price}, \citenamefont {Ozawa},\ and\ \citenamefont {Carusotto}}]{price2014quantum}%
  \BibitemOpen
  \bibfield  {author} {\bibinfo {author} {\bibfnamefont {H.~M.}\ \bibnamefont {Price}}, \bibinfo {author} {\bibfnamefont {T.}~\bibnamefont {Ozawa}},\ and\ \bibinfo {author} {\bibfnamefont {I.}~\bibnamefont {Carusotto}},\ }\bibfield  {title} {\bibinfo {title} {Quantum mechanics with a momentum-space artificial magnetic field},\ }\href@noop {} {\bibfield  {journal} {\bibinfo  {journal} {Phys. Rev. Lett.}\ }\textbf {\bibinfo {volume} {113}},\ \bibinfo {pages} {190403} (\bibinfo {year} {2014})}\BibitemShut {NoStop}%
\bibitem [{\citenamefont {Dong}\ \emph {et~al.}(2025)\citenamefont {Dong}, \citenamefont {Sommer}, \citenamefont {Soejima}, \citenamefont {Parker},\ and\ \citenamefont {Vishwanath}}]{dong2025phonons}%
  \BibitemOpen
  \bibfield  {author} {\bibinfo {author} {\bibfnamefont {J.}~\bibnamefont {Dong}}, \bibinfo {author} {\bibfnamefont {O.~E.}\ \bibnamefont {Sommer}}, \bibinfo {author} {\bibfnamefont {T.}~\bibnamefont {Soejima}}, \bibinfo {author} {\bibfnamefont {D.~E.}\ \bibnamefont {Parker}},\ and\ \bibinfo {author} {\bibfnamefont {A.}~\bibnamefont {Vishwanath}},\ }\bibfield  {title} {\bibinfo {title} {Phonons in electron crystals with berry curvature},\ }\href@noop {} {\bibfield  {journal} {\bibinfo  {journal} {arXiv preprint arXiv:2503.16390}\ } (\bibinfo {year} {2025})}\BibitemShut {NoStop}%
\bibitem [{\citenamefont {Onofri}(2001)}]{onofri2001landau}%
  \BibitemOpen
  \bibfield  {author} {\bibinfo {author} {\bibfnamefont {E.}~\bibnamefont {Onofri}},\ }\bibfield  {title} {\bibinfo {title} {Landau levels on a torus},\ }\href@noop {} {\bibfield  {journal} {\bibinfo  {journal} {International Journal of Theoretical Physics}\ }\textbf {\bibinfo {volume} {40}},\ \bibinfo {pages} {537} (\bibinfo {year} {2001})}\BibitemShut {NoStop}%
\bibitem [{\citenamefont {Fremling}(2015)}]{fremling2015quantum}%
  \BibitemOpen
  \bibfield  {author} {\bibinfo {author} {\bibfnamefont {M.}~\bibnamefont {Fremling}},\ }\emph {\bibinfo {title} {Quantum Hall wave functions on the torus}},\ \href@noop {} {Ph.D. thesis},\ \bibinfo  {school} {Department of Physics, Stockholm University} (\bibinfo {year} {2015})\BibitemShut {NoStop}%
\bibitem [{\citenamefont {Eisenstein}(2014)}]{eisenstein2014exciton}%
  \BibitemOpen
  \bibfield  {author} {\bibinfo {author} {\bibfnamefont {J.}~\bibnamefont {Eisenstein}},\ }\bibfield  {title} {\bibinfo {title} {Exciton condensation in bilayer quantum hall systems},\ }\href@noop {} {\bibfield  {journal} {\bibinfo  {journal} {Annu. Rev. Condens. Matter Phys.}\ }\textbf {\bibinfo {volume} {5}},\ \bibinfo {pages} {159} (\bibinfo {year} {2014})}\BibitemShut {NoStop}%
\bibitem [{\citenamefont {Verma}\ \emph {et~al.}(2024)\citenamefont {Verma}, \citenamefont {Guerci},\ and\ \citenamefont {Queiroz}}]{verma2024geometric}%
  \BibitemOpen
  \bibfield  {author} {\bibinfo {author} {\bibfnamefont {N.}~\bibnamefont {Verma}}, \bibinfo {author} {\bibfnamefont {D.}~\bibnamefont {Guerci}},\ and\ \bibinfo {author} {\bibfnamefont {R.}~\bibnamefont {Queiroz}},\ }\bibfield  {title} {\bibinfo {title} {Geometric stiffness in interlayer exciton condensates},\ }\href@noop {} {\bibfield  {journal} {\bibinfo  {journal} {Phys. Rev. Lett.}\ }\textbf {\bibinfo {volume} {132}},\ \bibinfo {pages} {236001} (\bibinfo {year} {2024})}\BibitemShut {NoStop}%
\bibitem [{\citenamefont {T{\"o}rm{\"a}}\ \emph {et~al.}(2018)\citenamefont {T{\"o}rm{\"a}}, \citenamefont {Liang},\ and\ \citenamefont {Peotta}}]{torma2018quantum}%
  \BibitemOpen
  \bibfield  {author} {\bibinfo {author} {\bibfnamefont {P.}~\bibnamefont {T{\"o}rm{\"a}}}, \bibinfo {author} {\bibfnamefont {L.}~\bibnamefont {Liang}},\ and\ \bibinfo {author} {\bibfnamefont {S.}~\bibnamefont {Peotta}},\ }\bibfield  {title} {\bibinfo {title} {Quantum metric and effective mass of a two-body bound state in a flat band},\ }\href@noop {} {\bibfield  {journal} {\bibinfo  {journal} {Phys. Rev. B}\ }\textbf {\bibinfo {volume} {98}},\ \bibinfo {pages} {220511} (\bibinfo {year} {2018})}\BibitemShut {NoStop}%
\bibitem [{\citenamefont {Cao}\ \emph {et~al.}(2018)\citenamefont {Cao}, \citenamefont {Fatemi}, \citenamefont {Fang}, \citenamefont {Watanabe}, \citenamefont {Taniguchi}, \citenamefont {Kaxiras},\ and\ \citenamefont {Jarillo-Herrero}}]{cao2018unconventional}%
  \BibitemOpen
  \bibfield  {author} {\bibinfo {author} {\bibfnamefont {Y.}~\bibnamefont {Cao}}, \bibinfo {author} {\bibfnamefont {V.}~\bibnamefont {Fatemi}}, \bibinfo {author} {\bibfnamefont {S.}~\bibnamefont {Fang}}, \bibinfo {author} {\bibfnamefont {K.}~\bibnamefont {Watanabe}}, \bibinfo {author} {\bibfnamefont {T.}~\bibnamefont {Taniguchi}}, \bibinfo {author} {\bibfnamefont {E.}~\bibnamefont {Kaxiras}},\ and\ \bibinfo {author} {\bibfnamefont {P.}~\bibnamefont {Jarillo-Herrero}},\ }\bibfield  {title} {\bibinfo {title} {Unconventional superconductivity in magic-angle graphene superlattices},\ }\href@noop {} {\bibfield  {journal} {\bibinfo  {journal} {Nature}\ }\textbf {\bibinfo {volume} {556}},\ \bibinfo {pages} {43} (\bibinfo {year} {2018})}\BibitemShut {NoStop}%
\bibitem [{\citenamefont {Han}\ \emph {et~al.}(2024)\citenamefont {Han}, \citenamefont {Lu}, \citenamefont {Hadjri}, \citenamefont {Shi}, \citenamefont {Wu}, \citenamefont {Xu}, \citenamefont {Yao}, \citenamefont {Cotten}, \citenamefont {Sedeh}, \citenamefont {Weldeyesus} \emph {et~al.}}]{han2024signatures}%
  \BibitemOpen
  \bibfield  {author} {\bibinfo {author} {\bibfnamefont {T.}~\bibnamefont {Han}}, \bibinfo {author} {\bibfnamefont {Z.}~\bibnamefont {Lu}}, \bibinfo {author} {\bibfnamefont {Z.}~\bibnamefont {Hadjri}}, \bibinfo {author} {\bibfnamefont {L.}~\bibnamefont {Shi}}, \bibinfo {author} {\bibfnamefont {Z.}~\bibnamefont {Wu}}, \bibinfo {author} {\bibfnamefont {W.}~\bibnamefont {Xu}}, \bibinfo {author} {\bibfnamefont {Y.}~\bibnamefont {Yao}}, \bibinfo {author} {\bibfnamefont {A.~A.}\ \bibnamefont {Cotten}}, \bibinfo {author} {\bibfnamefont {O.~S.}\ \bibnamefont {Sedeh}}, \bibinfo {author} {\bibfnamefont {H.}~\bibnamefont {Weldeyesus}}, \emph {et~al.},\ }\bibfield  {title} {\bibinfo {title} {Signatures of chiral superconductivity in rhombohedral graphene},\ }\href@noop {} {\bibfield  {journal} {\bibinfo  {journal} {arXiv preprint arXiv:2408.15233}\ } (\bibinfo {year} {2024})}\BibitemShut {NoStop}%
\bibitem [{\citenamefont {Xu}\ \emph {et~al.}(2025)\citenamefont {Xu}, \citenamefont {Sun}, \citenamefont {Li}, \citenamefont {Zheng}, \citenamefont {Xu}, \citenamefont {Gao}, \citenamefont {Jia}, \citenamefont {Watanabe}, \citenamefont {Taniguchi}, \citenamefont {Tong} \emph {et~al.}}]{xu2025signatures}%
  \BibitemOpen
  \bibfield  {author} {\bibinfo {author} {\bibfnamefont {F.}~\bibnamefont {Xu}}, \bibinfo {author} {\bibfnamefont {Z.}~\bibnamefont {Sun}}, \bibinfo {author} {\bibfnamefont {J.}~\bibnamefont {Li}}, \bibinfo {author} {\bibfnamefont {C.}~\bibnamefont {Zheng}}, \bibinfo {author} {\bibfnamefont {C.}~\bibnamefont {Xu}}, \bibinfo {author} {\bibfnamefont {J.}~\bibnamefont {Gao}}, \bibinfo {author} {\bibfnamefont {T.}~\bibnamefont {Jia}}, \bibinfo {author} {\bibfnamefont {K.}~\bibnamefont {Watanabe}}, \bibinfo {author} {\bibfnamefont {T.}~\bibnamefont {Taniguchi}}, \bibinfo {author} {\bibfnamefont {B.}~\bibnamefont {Tong}}, \emph {et~al.},\ }\bibfield  {title} {\bibinfo {title} {Signatures of unconventional superconductivity near reentrant and fractional quantum anomalous {Hall} insulators},\ }\href@noop {} {\bibfield  {journal} {\bibinfo  {journal} {arXiv preprint arXiv:2504.06972}\ } (\bibinfo {year} {2025})}\BibitemShut {NoStop}%
\bibitem [{\citenamefont {Ozawa}\ and\ \citenamefont {Mera}(2021)}]{ozawa2021relations}%
  \BibitemOpen
  \bibfield  {author} {\bibinfo {author} {\bibfnamefont {T.}~\bibnamefont {Ozawa}}\ and\ \bibinfo {author} {\bibfnamefont {B.}~\bibnamefont {Mera}},\ }\bibfield  {title} {\bibinfo {title} {Relations between topology and the quantum metric for {Chern} insulators},\ }\href@noop {} {\bibfield  {journal} {\bibinfo  {journal} {Phys. Rev. B}\ }\textbf {\bibinfo {volume} {104}},\ \bibinfo {pages} {045103} (\bibinfo {year} {2021})}\BibitemShut {NoStop}%
\end{thebibliography}

%

\end{document}